\documentclass[aps,prx,amsmath,floats,floatfix,twocolumn,
  superscriptaddress,nofootinbib,showpacs]{revtex4-1}

\pdfoutput=1 % Specifiy PDFLaTeX as per arXiv request. This has to be in the first 5 lines!
\usepackage{diagbox}
\usepackage{mathrsfs}
\usepackage{multirow}
\usepackage{braket}
\usepackage{amsfonts}
\usepackage{amsmath}
\usepackage{amssymb}
\usepackage{amsthm}
\usepackage{bm}
\usepackage{graphicx}
\usepackage{graphics}
\usepackage{latexsym}
\usepackage{rotating}
\usepackage[dvipsnames]{xcolor}
\usepackage{mathrsfs}
\usepackage{microtype}
\usepackage{verbatim}
\usepackage{revsymb}
\usepackage{url}

\usepackage[caption=false]{subfig}
\usepackage[normalem]{ulem}

\usepackage[breaklinks=true]{hyperref}
\hypersetup{
    colorlinks=true,
    linkcolor=NavyBlue,
    filecolor=Magenta,
    urlcolor=NavyBlue,
    citecolor=NavyBlue
}

\usepackage{yfonts}
\usepackage{float}
\usepackage{xspace}
\usepackage{mathrsfs}
\usepackage{siunitx}
\usepackage{array}

\allowdisplaybreaks[4]

\newcommand{\be}{\begin{equation}}
\newcommand{\ee}{\end{equation}}

\newcommand{\ba}{\begin{align}}
\newcommand{\ea}{\end{align}}

\makeatletter
\newcommand*{\rom}[1]{\expandafter\@slowromancap\romannumeral #1@}
\makeatother

\AtBeginDocument{%
    \newwrite\bibnotes
    \def\bibnotesext{Notes.bib}
    \immediate\openout\bibnotes=\jobname\bibnotesext
    \immediate\write\bibnotes{@CONTROL{REVTEX41Control}}
    \immediate\write\bibnotes{@CONTROL{%
    apsrev41Control,author="08",editor="1",pages="1",title="0",year="1"}}
    \if@filesw
    \immediate\write\@auxout{\string\citation{apsrev41Control}}%
    \fi
}
\allowdisplaybreaks

\begin{document}

\title{Spacetime of rotating black holes surrounded by massive scalar charges}

\author{Adrian Ka-Wai Chung}
\email{kwc43@cam.ac.uk}
\affiliation{DAMTP, Centre for Mathematical Sciences, University of Cambridge, Wilberforce Road, Cambridge CB3 0WA, United Kingdom}

\date{\today}

%%%%%%%%%%%%%%%%%%%%%%%%%%%%%%%%%%%%%%%%%%%%%%%%%%%%%%%%%%%%%%
\begin{abstract} 
Massive scalar charges are ubiquitous in extensions to General Relativity and the Standard Model in particle physics. 
We describe spectral methods which can accurately construct the spacetime of rotating black holes with dimensionless spin up to $a \leq 0.8$ surrounded by massive scalar fields nonminimally coupled to spacetime curvature.
We consider axi-dilaton, dynamical Chern–Simons, and scalar Gauss–Bonnet couplings, and obtain leading-order solutions for both the scalar field and the associated metric modifications.
Our method accurately resolves massive scalar fields with Compton wavelengths as short as 5 times the black-hole mass, achieving residual errors $\lesssim 10^{-5}$, and yields the corresponding leading-order spacetime modifications with residual errors $\lesssim 10^{-3}$.
Using the constructed spacetimes, we computes the leading-order shifts in the surface gravity and the angular velocity of the event horizon, important information for computing the quasinormal modes.
These results pave the way to incorporate massive scalar charges into electromagnetic observations and gravitational-wave detections of black holes, potentially enabling new probes of fundamental scalar degrees of freedom.
\end{abstract}

\maketitle

\vspace{0.2cm}
\noindent 

%%%%%%%%%%%%%%%%%%%%%%%%%%%%%%%%%%%%%%%%%%%%%%%%%%%%%%%%%%%%%%%%%%%%%%%%%%%%%%%%%%%%%%%%%%%%%%%%%%%%%%%%%%%%%%%%%%%%%%%%%%%%
\section{Introduction}

Some astrophysical phenomena, such as the rotation curves of galaxies~\cite{rotation_curve_01,rotation_curve_02}, gravitational lensing by galactic and cluster-scale structures~\cite{Massey:2010hh,Gavazzi:2007vw,Guzik:2002zp,Mandelbaum:2005nx}, the late-time accelerated expansion of the Universe~\cite{late_time_acceleration_01,late_time_acceleration_02}, and the observed baryon asymmetry~\cite{Farrar:1993sp}, cannot be fully addressed by General Relativity without invoking the concepts of dark energy or dark matter. 
This inability points to two open questions in fundamental physics: whether General Relativity remains the correct description of gravity in the strong-field regime, and what the true nature of dark matter and dark energy is.

The first open question has motivated the proposals of extensions to General Relativity. 
Such extensions often introduce additional scalar degrees of freedom that couple nonminimally to curvature through higher-degree/order curvature quantities. 
Examples of these theories include axi-dilaton \cite{Kallosh:2022vha, Cano:2021rey}, dynamical Chern-Simons \cite{dCS_02, Alexander:2009tp, Yagi:2015oca, Nair:2019iur, Perkins:2021mhb}, and scalar Gauss-Bonnet \cite{Ripley:2019irj, Julie:2024fwy, Julie:2019sab, Damour:1992we} gravity. 
These theories could be regarded as an effective-field theory approximation to grand unification frameworks, such as string theory, or quantum-gravity candidates, such as loop-quantum gravity, in the low-energy limit \cite{Cano_Ruiperez_2019, GROSS19861, Gross:1985rr, Alexander:2009tp}. 
Apart from addressing the two open questions, the axi-dilaton theory provides a possible mechanism of cosmological inflation \cite{Kallosh:2022vha}, and dynamical Chern-Simons theory might also address the baryon asymmetry in the observable Universe \cite{dCS_03, Alexander:2004us}. 
These extensions deform black-hole spacetimes from the ones in General Relativity \cite{Kanti:1995vq, Yunes:2009hc, Pani:2009wy, Sullivan:2019vyi, Sullivan:2020zpf, Cano_Ruiperez_2019, Lam:2025elw, Lam:2025fzi}. 
The deformations could leave signatures in electromagnetic or gravitational waves generated by astrophysical processes involving black holes, making black-hole potent laboratories to test these extensions.

Setting aside their nonminimal couplings to spacetime curvature, additional scalar fields, particularly those with nonzero mass, are themselves well-motivated dark-matter candidates~\cite{Hu:2000ke,Marsh:2015xka,Hui:2016ltb}.
One intriguing astrophysical consequence of massive scalar fields is black-hole superradiance: a process by which massive scalar particles or dispersive waves can extract rotational energy from a spinning black hole when their frequencies satisfy the superradiant condition~\cite{Starobinskii:1973vzb,Brito:2015oca}.
Through this mechanism, massive scalar fields can form macroscopic condensates around black holes~\cite{Brito:2015oca}, thereby modifying their astrophysical properties and potentially leaving observable imprints in gravitational-wave and electromagnetic signals~\cite{Cardoso:2011xi, Yunes:2011aa, Brito:2014wla, Brito:2017zvb, Maselli:2020zgv,Maselli:2021men,Berti:2019wnn,Zhang:2018kib,Ferreira:2017pth, Psaltis:2018xkc,Baumann:2018vus,EventHorizonTelescope:2022xqj,Uniyal:2025uvc}.
Black-hole superradiance can also occur in modified gravity theories, including the aforementioned theories~\cite{Alexander:2022avt}.
In such theories, nonminimal couplings to spacetime curvature allow massive scalar charges to build stationary configurations around black holes.
These configurations lead to signatures that render the massive scalar charges directly searchable in existing gravitational-wave data.
Moreover, such stationary configurations may serve as natural initial conditions for triggering superradiant instabilities.

To test General Relativity and search for dark-matter candidates using black holes, it is essential to understand the spacetime of rotating black holes with nonminimal coupling to spacetime curvature and with dark matter.
However, constructing such spacetimes presents significant theoretical challenges, as it requires solving a system of coupled, nonlinear partial differential equations.
Moreover, astrophysical black holes are often rapidly rotating, which further complicates the problem.

Nevertheless, the problem can be simplified by exploiting well-motivated approximations.
Current observational and theoretical constraints on extensions to General Relativity suggest that any such couplings, if present, must be small.
This motivates focusing attention to the scalar-field configuration and the associated spacetime modifications at leading order in the coupling parameter.
Within this small-coupling regime, the otherwise nonlinear field equations governing spacetime deformations reduce to a linear system defined on a fixed general-relativistic background, substantially simplifying the analysis.
For massless scalar fields, the leading-order scalar profiles and corresponding spacetime modifications have recently been computed accurately using spectral methods~\cite{Lam:2025elw,Lam:2025fzi}.

The goal of this paper is to extend the spectral methods in~\cite{Lam:2025elw,Lam:2025fzi} to construct the spacetime modifications of rotating black holes surrounded by massive scalar charges.
In particular, we apply these extended spectral methods to construct spacetimes of rotating black holes sourced by massive scalar fields in axi-dilaton, dynamical Chern–Simons, and scalar Gauss–Bonnet gravity, motivated by their broad relevance in fundamental physics.
The paper is organized as follows.
In Sec.~\ref{sec:model}, we introduce a Lagrangian density that encompasses all three theories and define the relevant coupling constants, parameters, and scalar-field masses.
From this Lagrangian, we derive the equations of motion governing the scalar field (the Klein–Gordon equation) and the spacetime geometry (the modified Einstein equations).
By analyzing these equations, we determine the asymptotic behavior of the massive scalar field and the spacetime modifications at the event horizon and at spatial infinity.
These asymptotic properties are then used to construct suitable \textit{ansätze} for the massive scalar field in Sec.~\ref{sec:massive_scalar_charges} and for the metric deformations in Sec.~\ref{sec:Metric_ansatz}, working in Boyer–Lindquist coordinates in which the radial location of the event horizon remains unchanged.
In Sec.~\ref{sec:Scalar_field}, we develop spectral methods to construct massive scalar-field solutions accurate to leading order in the coupling parameter.
We introduce several diagnostics to quantify the convergence and accuracy of the scalar-field solutions.
We find that both the backward modulus difference and the scalar-field residual decrease exponentially with spectral order up to a characteristic value, beyond which the convergence continues at a slower rate.
For scalar fields with Compton wavelengths shorter than approximately 10 times the black-hole mass, both measures reach a minimum at spectral orders $N \lesssim 30$.
Based on these diagnostics, we define a systematic procedure for selecting the most accurate scalar-field solution.
Importantly, we find that while increasing the scalar-field mass leads to a more rapid radial decay of the field, it does not significantly alter its multipolar structure.
In Sec.~\ref{sec:Metric}, we construct the spacetime modifications of the host black hole using the optimal scalar-field solutions.
The metric modifications exhibit convergence and accuracy properties similar to those of the scalar field.
As in the scalar-field case, we find that the scalar-field mass does not significantly affect the geometric structure of the spacetime modifications, although it does modify their overall magnitude.
Our results enhance the understanding of spacetimes of rotating black holes surrounded by matter. 
In Sec.~\ref{sec:Physical_properties}, we compute the surface gravity and the angular velocity of the event horizon.
These physical properties are the crucial ingredients for the computation of quasinormal-mode spectra \cite{Chung:2024vaf, Chung:2025gyg}, which could be used to search for the massive scalar charges via black-hole ringdown spectroscopy. 
Finally, in Sec.~\ref{sec:Conclusions}, we discuss the astrophysical implications of our results and outline directions for future work.

Throughout the paper, we adopt the following conventions.
We use coordinates $x^{\mu}=(t,r,\chi,\phi)$, where $\chi=\cos\theta$ and $\theta$ is the polar angle.
The metric signature is $(-,+,+,+)$, and we work in geometric units with $c=G=M=1$, where $M$ is the mass of the black hole, which has been normalized to unitary.
Following the convention of Ref.~\cite{Lam:2025fzi}, the leading-order correction in the small-coupling parameter $\zeta$ [defined in Eq.~(\ref{eq:coupling_parameter})] to any quantity $Q$ is written as
\begin{equation}
Q = Q^{(0)} + \zeta Q^{(1)},
\end{equation}
where $Q^{(0)}$ denotes the corresponding general-relativistic quantity and $Q^{(1)}$ its leading-order correction.
All scalar-field and spacetime-modification solutions constructed in this work are provided as Supplementary Material.

%%%%%%%%%%%%%%%%%%%%%%%%%%%%%%%%%%%%%%%%%%%%%%%%%%%%%%%%%%%%%%%%%%%%%%%%%%%%%%%%%%%%%%%%%%%%%%%%%%%%%%%%%%%%%%%%%%%%%%%%%%%%
\section{Models} \label{sec:model}

\subsection{Lagrangian density}

The Lagrangian density of the theories considered in this work can be written compactly as~\cite{Cano_Ruiperez_2019}
\begin{equation}\label{eq:Lagrangian}
\begin{split}
16 \pi \mathscr{L} = & R + \ell_1{}^2 \vartheta_1 \mathscr{G} + \ell_2{}^2 \left( \vartheta_1 \sin \theta_m + \vartheta_2 \cos \theta_m \right) \mathscr{P} \\
& - \frac{1}{2} \nabla_{\nu} \vartheta_1 \nabla^{\nu} \vartheta_1  - \frac{1}{2} \mu_1 {}^2 \vartheta_1 {}^2 \\
& - \frac{1}{2} \nabla_{\nu} \vartheta_2 \nabla^{\nu} \vartheta_2 - \frac{1}{2} \mu_2 {}^2 \vartheta_2 {}^2.
\end{split}
\end{equation}
Here $\ell_{1,2}$ are the quadratic-gravity coupling length scales, each with dimensions of length.
The fields $\vartheta_1$ and $\vartheta_2$ denote a scalar and a pseudoscalar field, respectively, which couple nonminimally to the spacetime curvature.
$\mathscr{G} = R^{2} - 4 R_{\alpha\beta}R^{\alpha\beta} + R_{\alpha\beta\gamma\delta}R^{\alpha\beta\gamma\delta}$~\cite{Ripley:2019irj} is the Gauss–Bonnet invariant, while $\mathscr{P} = R_{\mu \nu \rho \sigma}{ } \tilde{R}^{\mu \nu \rho \sigma}$ is the Pontryagin invariant~\cite{Alexander:2009tp}.
Although $\vartheta_2$ is a pseudoscalar field, we shall collectively refer to $\vartheta_{1,2}$ by the ``scalar field/charges". 
The parameter $\theta_m$ is a mixing angle that controls the relative contribution of the Gauss–Bonnet and Pontryagin terms in the Lagrangian.
Several well-known theories are recovered as limiting cases of Eq.~\eqref{eq:Lagrangian}.
When $\ell_2=0$, the mixing angle $\theta_m$ becomes irrelevant and the Lagrangian reduces to that of scalar Gauss–Bonnet gravity.
When $\ell_1=0$ and $\theta_m=0$, the theory reduces to dynamical Chern–Simons gravity.
When $\ell_1=\ell_2\neq0$ and $\theta_m=0$, the Lagrangian corresponds to axi-dilaton gravity.
The parameters $\mu_{1,2}$ denote the masses of the scalar fields $\vartheta_{1,2}$ and have dimensions of inverse length.
For $\mu_{q}>0$, the theory describes gravity coupled to massive scalar field(s), with Compton wavelengths given by $\mu_{q}{}^{-1}$.

Following Refs.~\cite{Chung:2024vaf, Chung:2025gyg, Lam:2025elw, Lam:2025fzi}, we introduce a rescaled scalar field $\vartheta_{q} = \ell_{q} {}^{2} \bar{\vartheta}_{q}$.
Varying the Lagrangian density in Eq.~\eqref{eq:Lagrangian} with respect to the metric tensor and the scalar field(s), for given $\ell_{1,2}$ and mixing angle $\theta_m$, yields the equations of motion (with no implied summation over $q=1,2$),
\begin{align}
& R_{\beta}{}^{\nu} + \lambda^4 \zeta_q [(\mathscr{A}_q)_{\beta}{}^{\nu} - (\bar{T}_q)_{\beta}{}^{\nu}] = 0, \label{eq:MEE} \\
& \Box \bar{\vartheta}_q - \mu_q {}^2 ~\bar{\vartheta}_q + \mathscr{Q}_q = 0, \label{eq:SFE}
\end{align}
Here $\zeta_q$ is a dimensionless coupling parameter defined as
\begin{equation}\label{eq:coupling_parameter}
\zeta_q = \left( \frac{\ell_q}{\lambda} \right)^4, 
\end{equation}
where $\lambda$ is a characteristic length scale of the problem.
Since our goal is to construct stationary massive scalar-field configurations around black holes and to determine their leading-order effects on the surrounding spacetime, a natural choice is $\lambda=M$, where $M$ is the black-hole mass.
As, throughout this paper, the unit of $M=1$ has been enforced, $\lambda = M = 1$. 
In this unit, the factor $\lambda^4$ in Eq.~\eqref{eq:MEE} reduces to unity, and the field equation simplifies to
\begin{equation}\label{eq:MEE2}
R_{\beta}{}^{\nu} + \zeta_q [(\mathscr{A}_q)_{\beta}{}^{\nu} - (\bar{T}_q)_{\beta}{}^{\nu}] = 0.
\end{equation}
The tensor $(\mathscr{A}_q)_{\mu}{}^{\nu}$ is a rank-$(1,1)$ tensor constructed from curvature tensors and derivatives of the scalar field $\bar{\vartheta}_q$.
For sGB gravity, $(\mathscr{A}_{\rm sGB})_{\mu}{}^{\nu}$ is given by \cite{East:2020hgw, Cano_Ruiperez_2019, Lam:2025fzi} 
\begin{equation}\label{eq:AuusGB}
\begin{split}
    (\mathscr{A}_{\rm sGB})_{\mu}{}^{\nu} &= \left[\delta_{\mu \lambda \gamma \delta}^{\nu \sigma \alpha \beta} - \frac{1}{2} \delta_{\mu}{}^{\nu}\delta_{\eta \lambda \gamma \delta}^{\eta \sigma \alpha \beta}\right] R^{\gamma \delta}{}_{\alpha \beta} \nabla^{\lambda} \nabla_{\sigma} \bar{\vartheta}_{\rm sGB}, 
\end{split}
\end{equation}
where $\delta^{\nu \sigma \alpha \beta}_{\mu \lambda \gamma \delta}$ is the generalized Kronecker delta, defined as 
\begin{equation}
\delta_{\mu_1 \mu_2 \mu_3 \mu_4}^{\nu_1 \nu_2 \nu_3 \nu_4} = \det 
\begin{pmatrix}
\delta_{\mu_1}^{\nu_1} & \delta_{\mu_2}^{\nu_1} & \delta_{\mu_3}^{\nu_1} & \delta_{\mu_4}^{\nu_1} \\
\delta_{\mu_1}^{\nu_2} & \delta_{\mu_2}^{\nu_2} & \delta_{\mu_3}^{\nu_2} & \delta_{\mu_4}^{\nu_2} \\
\delta_{\mu_1}^{\nu_3} & \delta_{\mu_2}^{\nu_3} & \delta_{\mu_3}^{\nu_3} & \delta_{\mu_4}^{\nu_3} \\
\delta_{\mu_1}^{\nu_4} & \delta_{\mu_2}^{\nu_4} & \delta_{\mu_3}^{\nu_4} & \delta_{\mu_4}^{\nu_4} \\
\end{pmatrix}. 
\end{equation}
For dCS gravity, the source term in the scalar-field equation is $\mathscr{Q}=\mathscr{P}$, $(\mathscr{A}_{\rm dCS})_{\mu}{}^{\nu} = g_{\mu \alpha} (\mathscr{A}_{\rm dCS})^{\alpha \nu}$, where $(\mathscr{A}_{\rm dCS})^{\mu \nu}$ is defined as \cite{Yunes:2009hc}
\begin{equation}\label{eq:AuudCS}
\begin{split}
    (\mathscr{A}_{\rm dCS})^{\mu\nu} &= -4\Big[\left(\nabla_\sigma \bar{\vartheta}_{\rm dCS} \right) \varepsilon^{\sigma \delta \alpha(\mu|} \nabla_\alpha R^{|\nu)}{}_{\delta} \\
    &\qquad\qquad + \left(\nabla_\sigma \nabla_\delta \bar{\vartheta}_{\rm dCS} \right) \tilde{R}^{\delta (\mu \nu) \sigma}\Big].
\end{split}
\end{equation}
$(\bar{T}_q)_{\mu}{}^{\nu}$ denotes the trace-reversed stress–energy tensor of the scalar field, given by \cite{Alexander:2022avt}
\begin{align}\label{eq:Energy_momentum_tensor}
(\bar T_q)_{\beta}{}^{\nu}
= \frac{1}{2}\nabla_{\beta}\bar{\vartheta}_q \, \nabla^{\nu}\bar{\vartheta}_q
+ \frac{1}{4}\delta_{\beta}{}^{\nu}\,\mu_q{}^{2}\,\bar{\vartheta}_q {}^{2}.
\end{align}

The existing constraints on beyond-Einstein couplings indicate that the dimensionless parameter $\zeta$, if nonzero, must be small~\cite{Chung:2025wbg,Xie:2024xex,Julie:2024fwy}. 
Thus, we solve Eqs.~\eqref{eq:MEE} and \eqref{eq:SFE} using a perturbative expansion in $\zeta$.
Specifically, we expand the metric tensor and scalar field as power series in $\zeta$ \footnote{
We note that, due to the rescaling $\vartheta = \ell^2 \bar{\vartheta}$, the leading nontrivial scalar configuration appears at zeroth order in our expansion. In conventional effective-field-theory counting, this corresponds to the leading-order [$\mathcal{O}(\ell^2)$] scalar field. This choice of normalization simplifies the perturbative structure of the field equations.
}, 
\begin{align}
g_{\mu\nu} &= g_{\mu\nu}^{(0)} + \zeta g_{\mu\nu}^{(1)} + {\cal{O}}(\zeta^2), \\
\bar{\vartheta} &= \bar{\vartheta}^{(0)} + \zeta \bar{\vartheta}^{(1)} + {\cal{O}}(\zeta^2). 
\end{align}
Substituting these expansions into Eqs.~\eqref{eq:MEE} and \eqref{eq:SFE}, we find that at the zeroth order, 
\begin{align}
[R_{\beta}{}^{\nu}]^{(0)} &= 0, \label{eq:EE0}
\end{align}
\begin{align}
E_{\vartheta} := \Box^{(0)} \bar{\vartheta}_q - \mu_q {}^2 ~\bar{\vartheta}_q + \mathscr{Q}_q {}^{(0)} &= 0. \label{eq:SF0}
\end{align}
Here $[R_{\beta}{}^{\nu}]^{(0)}$ denotes the Ricci tensor evaluated on the background metric $g_{\mu\nu}^{(0)}$, while $\Box^{(0)}$ and $\mathscr{Q}_q{}^{(0)}$ are, respectively, the d’Alembertian operator and the quadratic curvature invariant computed with respect to the same background.
Since the beyond-Einstein coupling and the massive scalar field are expected to deform the general-relativistic solution only at order $\mathcal{O}(\zeta)$, we take the background metric $g_{\mu\nu}^{(0)}$ to be the Kerr spacetime throughout this work.

Inspecting Eqs.~\eqref{eq:MEE} and \eqref{eq:SFE}, we observe that, in order to solve for the first-order metric perturbation $g_{\mu\nu}^{(1)}$, it suffices to evaluate $(\mathscr{A}_q)_{\mu}{}^{\nu}$ and $(\bar{T}_q)_{\mu}{}^{\nu}$ using the background metric $g_{\mu\nu}^{(0)}$ and the leading-order scalar field $\bar{\vartheta}_q^{(0)}$ in Eq.~\eqref{eq:MEE}.
Schematically, the equations governing $g_{\mu\nu}^{(1)}$ can be written as
\begin{align}\label{eq:EE1}
E_{\beta}{}^{\nu} := [R_{\beta}{}^{\nu}]^{(1)} + [\mathscr{A}_{\beta}{}^{\nu}]^{(0)} - [\bar{T}_{\beta}{}^{\nu}]^{(0)} = 0. 
\end{align}
Here $\bigl[R_{\beta}{}^{\nu}\bigr]^{(1)}$ denotes the $\mathcal{O}(\zeta)$ correction to the Ricci tensor computed from the deformed metric $g_{\mu\nu}=g_{\mu\nu}^{(0)}+\zeta g_{\mu\nu}^{(1)}$.
This term depends on the background metric $g_{\mu\nu}^{(0)}$ and linearly on the metric modifications $g_{\mu\nu}^{(1)}$ and their derivatives.
By contrast, $\bigl[\mathscr{A}_{\beta}{}^{\nu}\bigr]^{(0)}$ and $\bigl[\bar{T}_{\beta}{}^{\nu}\bigr]^{(0)}$ are obtained by evaluating $(\mathscr{A}_q)_{\beta}{}^{\nu}$ and $(\bar{T}_q)_{\beta}{}^{\nu}$ with $g_{\mu\nu}^{(0)}$ and the leading-order scalar field $\bar{\vartheta}_q^{(0)}$, and therefore do not depend on $g_{\mu\nu}^{(1)}$.
Using the background field equation Eq.~(\ref{eq:EE0}), $\left[(\mathscr{A}_{\rm dCS})^{\mu\nu} \right]^{(0)}$ can be simplified as 
\begin{equation}\label{eq:AuudCS}
\begin{split}
    \left[(\mathscr{A}_{\rm dCS})^{\mu\nu} \right]^{(0)} &= -4 \left[ \left(\nabla_\sigma \nabla_\delta \bar{\vartheta}_{\rm dCS} \right) \tilde{R}^{\delta (\mu \nu) \sigma} \right]^{(0)}.
\end{split}
\end{equation}

\subsection{Massive scalar charges}
\label{sec:massive_scalar_charges}

In this subsection, we examine the scalar-field equation $E_{\vartheta}=0$ in greater detail to determine the asymptotic behavior of the massive scalar field.
This analysis provides essential guidance for constructing an appropriate ansatz for the scalar field in the implementation of our spectral methods.
Since we seek stationary and axisymmetric solutions for $\bar{\vartheta}$, the scalar-field equation takes the explicit form
\begin{align}\label{eq:SF0-explicit}
\frac{1}{\Sigma}\frac{\partial}{\partial r}\left(\Delta \frac{\partial \bar{\vartheta}}{\partial r}\right) + \frac{1}{\Sigma} \frac{\partial}{\partial \chi}\left[(1-\chi^2)\frac{\partial \bar{\vartheta}}{\partial \chi}\right] - \mu^2 \bar{\vartheta}
&= - \mathscr{Q}^{(0)}\,, 
\end{align}
where, for simplicity, we have suppressed the field index $q$.

At spatial infinity, the source term behaves as $\mathscr{Q}^{(0)}\sim r^{-6}$.
Consequently, the Klein–Gordon equation asymptotically reduces to its homogeneous form, which in the far-field limit reads\footnote{
This is also the asymptotic form of the Klein–Gordon equation governing an ultralight boson with zero frequency at spatial infinity.
In that context, one may further infer that the scalar field behaves as $\vartheta \sim r^{-1-\mu} e^{-\mu r}$~\cite{Dolan:2007mj}.
However, we find that explicitly including the additional factor of $r^{-\mu}$ does not provide a noticeable improvement in numerical accuracy or efficiency.
For this reason, we do not incorporate this factor into the asymptotic ansatz for the scalar field.
}
\begin{align}
\nabla^2 \bar{\vartheta} - \mu^2 \bar{\vartheta} = 0, 
\end{align}
where $\nabla^{2}$ denotes the flat-space Laplacian expressed in spherical coordinates.
The leading-order asymptotic solution to this equation is well known~\cite{Feynman:1963uxa} and takes the form, 
\begin{equation}
\bar{\vartheta} (r \rightarrow \infty, \text{given }\chi) \sim \frac{e^{-\mu r}}{r} \left( 1 + \frac{A}{r} + \frac{B}{r^2} + .... \right), 
\end{equation}
where $A$ and $B$ are constants determined by subleading corrections.
Motivated by this asymptotic structure, we factor out the exponential decay and write
\begin{equation}\label{eq:varphi_def}
\bar{\vartheta} = e^{-\mu r} \varphi, 
\end{equation}
where the auxiliary function $\varphi$ vanishes as $r\to\infty$ and admits an expansion in inverse powers of $r$.

Near the event horizon, we can write Eq.~\eqref{eq:SF0-explicit} in the form, 
\begin{equation}
\begin{split}
& \frac{\partial^2 \bar{\vartheta}}{\partial r^2} + \frac{2r-r_{-}-r_{+}}{\Delta} \frac{\partial \bar{\vartheta}}{\partial r} - \mu^2 \frac{r^2+\chi^2}{\Delta} \bar{\vartheta}  \\
& = \frac{\Sigma \mathscr{Q}}{\Delta} - \frac{1}{\Delta} \frac{\partial}{\partial \chi} \left( (1-\chi^2) \frac{\partial \bar{\vartheta}}{\partial \chi} \right)
\end{split}
\end{equation}
Any physical solution for $\bar{\vartheta}$ must possess finite derivatives with respect to $\chi$ at the event horizon.
For a fixed angular position $\chi=\chi_{0}$, we expand the equation in the limit $r\to r_{+}$, the leading-order behavior reduces to
\begin{equation}
\frac{\partial^2 \bar{\vartheta}}{\partial r^2} + \frac{1}{r-r_+} \frac{\partial \bar{\vartheta}}{\partial r} + \frac{\alpha}{r-r_+} \bar{\vartheta} - \frac{\beta}{r-r_+} = 0, 
\end{equation}
where 
\begin{equation}
\begin{split}
\alpha = & \mu^2 \frac{r_+^2 + \chi_0{}^2}{r_+-r_-} \\
\beta = & \frac{\left[\Sigma \mathscr{Q}\right](r_+, \chi_0)}{r_+-r_-} - \frac{1}{r_+-r_-} \frac{\partial}{\partial \chi} \left( (1-\chi^2) \frac{\partial \bar{\vartheta}}{\partial \chi} \right) \Bigg|_{\chi_0}. 
\end{split}
\end{equation}
Here $\left[\Sigma \mathscr{Q}\right](r_+, \chi_0)$ denotes the product of $\Sigma $ and $\mathscr{Q}$ evaluated at the event horizon and at $\chi=\chi_{0}$.
The complementary (homogeneous) solution of this equation is given by
\begin{equation}
\begin{split}
\bar{\vartheta} \sim C_1 J_0 \left[2 \sqrt{\alpha (r-r_+)} \right] + C_2 Y_0 \left[2 \sqrt{\alpha (r-r_+)} \right], 
\end{split}
\end{equation}
where $C_{1}$ and $C_{2}$ are integration constants, and $J_{0}$ and $Y_{0}$ are Bessel functions of the first and second kind of the zeroth order.
Since $Y_0$ diverges as the argument vanishes, physical solutions demand $C_1 = 0$, which leads to finite solution at the event horizon. \footnote{A shorter, albeit less rigorous, argument is that physical solutions must be finite at the horizon.
Since $\varphi=e^{\mu r_{+}}\bar{\vartheta}$ differs from $\bar{\vartheta}$ only by a constant factor at $r=r_{+}$, regularity of $\bar{\vartheta}$ immediately implies regularity of $\varphi$.}
This asymptotic analysis therefore demonstrates that the auxiliary field $\varphi$, defined in Eq.~\eqref{eq:varphi_def}, remains finite and differentiable throughout the entire computational domain.
As a result, $\varphi$ admits a well-behaved spectral representation, justifying the use of spectral methods in our construction.

\subsection{Rotating black-hole spacetime}
\label{sec:Metric_ansatz}

Following Refs.~\cite{Cano_Ruiperez_2019, Lam:2025elw, Lam:2025fzi}, we adopt the following metric ansatz for a rotating black hole surrounded by massive scalar charges:
\begin{widetext}
\begin{equation}\label{eq:metric}
\begin{split}
ds^2 &= g_{\mu \nu}^{(0)} dx^{\mu}  dx^{\nu} \\
& = - \left( 1-\frac{2 M r}{\Sigma} - \zeta H_1(r, \chi)\right) dt^2 - \left[ 1 + \zeta H_2(r, \chi) \right] \frac{4 M^2 a r}{\Sigma} (1 - \chi^2) d \phi dt \\
& \quad + \left[ 1 + \zeta H_3(r, \chi) \right] \left( \frac{\Sigma}{\Delta} dr^2 + \frac{\Sigma}{1 - \chi^2} d \chi^2 \right) + \left[ 1 + \zeta H_4(r, \chi) \right](1-\chi^2) \left[r^{2} + M^2 a^{2}+\frac{2 M^3 a^{2} r}{\Sigma} (1 - \chi^2)\right] d\phi^2, \\
\end{split}
\end{equation}
\end{widetext}
where $H_i(r,\chi)$ denote the leading-order-in-$\zeta$ metric deformations induced by the beyond-Einstein coupling mediated by the massive scalar field.
To preserve asymptotic flatness and to ensure that $M$ and $J=M^2 a$ retain their interpretations as the Arnowitt–Deser–Misner (ADM) mass and angular momentum of the black hole, respectively, we impose the following boundary conditions~\cite{Cano_Ruiperez_2019, Lam:2025elw, Lam:2025fzi}: 
\begin{equation}\label{eq:BoundaryCondition}
    H_1^{(0)} = 0, \quad H_2^{(0)} = \frac{H_3^{(1)}}{2M}, \quad H_3^{(0)} = H_4^{(0)} = -\frac{H_3^{(1)}}{M}.  
\end{equation}
Here the functions $H_i^{(0)}(\chi)$ and $H_i^{(1)}(\chi)$ are defined through the asymptotic expansion, 
\begin{align} 
H_i(r, \chi) = H_i^{(0)} + \frac{1}{r} H_i^{(1)} + {\cal{O}}(r^{-2})\,. 
\end{align}

%%%%%%%%%%%%%%%%%%%%%%%%%%%%%%%%%%%%%%%%%%%%%%%%%%%%%%%%%%%%%%%%%%%%%%%%%%%%%%%%%%%%%%%%%%%%%%%%%%%%%%%%%%%%%%%%%%%%%%%%%%%%
\section{Construction of the massive scalar fields}\label{sec:Scalar_field}

Although the Klein–Gordon equation governing massive scalar fields closely resembles its massless counterpart ($\mu=0$), obtaining analytic solutions in the massive case is substantially more challenging.
In particular, even after performing a spin expansion, the solutions are no longer simple polynomials in $r^{-1}$ or in $\chi$, but instead of more complicated analytical form. \footnote{We are in debt to Pablo Cano for pointing this out.}
This difficulty motivates the development of spectral methods to construct accurate approximate solutions to the Klein–Gordon equation for massive scalar fields, which is the goal of this section.
We begin by outlining the analytical framework underlying our spectral approach to solving the Klein–Gordon equation.
We then describe the details of the numerical implementation.
Finally, we present and analyze the numerical results obtained using these spectral methods.

\subsection{Spectral methods}

Although $\bar{\vartheta}_q$ is finite and continuous over the domain $r\in(r{+},\infty)$ and $\chi\in[-1,1]$, a direct spectral expansion of $\bar{\vartheta}_q$ is prone to numerical instability.
In fact, expanding $\bar{\vartheta}_q$ as a linear combination of spectral basis functions leads to poor convergence, particularly for larger scalar-field masses.
The source of this instability is the exponentially decaying asymptotic behavior, $e^{-\mu r}/r$, which is challenging to resolve accurately with a finite number of spectral modes.
To accommodate this asymptotic structure and improve numerical stability, we instead perform the spectral expansion on an auxiliary field $\varphi$, defined by
\begin{equation}
\bar{\vartheta} = e^{-\mu r} \varphi, 
\end{equation}
so that $\varphi$ vanishes smoothly as $r\to\infty$.
This redefinition isolates the rapid exponential decay analytically, leaving $\varphi$ as a smoother function that admits a more stable spectral representation.

Note that Eq.~\eqref{eq:SF0-explicit} involves only rational functions of $r$ and $\chi$.
When the d’Alembertian operator acts on $\varphi$, the resulting equation contains coefficients multiplying derivatives of $\varphi$ that are products of the exponential factor $e^{-\mu r}$ and rational functions of $r$ and $\chi$.
Schematically, this equation can be written as
\begin{equation}\label{eq:KG_0}
e^{- \mu r} \left( \sum_{k, l} \mathscr{C}_{kl}(r, \chi) \partial_{r}^k \partial_{\chi}^l \varphi \right) = - \mathscr{Q}, 
\end{equation}
where $\mathscr{C}_{kl}(r,\chi)$ are rational functions of $r$ and $\chi$.
To cast this equation into a form suitable for spectral methods, we extract a common denominator by factorizing the expression $\sum_{k, l} \mathscr{C}_{kl}(r, \chi) \partial_{r}^k \partial_{\chi}^l \varphi + \mathscr{Q} $ (without including the factor $e^{- \mu r}$). 
Multiplying both sides of Eq.~\eqref{eq:KG_0} by this common denominator yields
\begin{equation}\label{eq:KG_r}
e^{- \mu r} \left( \sum_{i, j, k, l} d_{i,j,k,l} r^i \chi^j \partial_{r}^k \partial_{\chi}^l \varphi \right) = \sum_{i,j} q_{i,j} r^i \chi^j, 
\end{equation}
where the coefficients $d_{i,j,k,l}$ depend on $M$, $a$, and $\mu$, while $q_{i,j}$ depend only on $M$ and $a$.
For brevity, we have suppressed the field index $q$.
We emphasize that retaining the exponential factor $e^{-\mu r}$ explicitly on the derivative side of the equation is a crucial ingredient of our spectral scheme, as it isolates the stiff radial dependence analytically and significantly improves numerical stability.

As the equation in this form is defined on the semi-infinite radial domain $r\in(r_{+},\infty)$, we introduce a compactified radial coordinate $z$, defined by
\begin{equation}
z = \frac{2r_{+}}{r}-1,
\end{equation}
to facilitate the implementation of spectral methods.
Under this transformation, the event horizon $r=r_{+}$ is mapped to $z=+1$, while spatial infinity $r\to\infty$ is mapped to $z=-1$.
Since $z$ is a rational function of $r$, $\sum_{i, j, k, l} d_{i,j,k,l} r^i \chi^j \partial_{r}^k \partial_{\chi}^l \varphi$ and $\sum_{i,j} q_{i,j} r^i \chi^j$ will also consist of only rational function of $z$ and $\chi$. 
Following the same procedure used to obtain Eq.~\eqref{eq:KG_r}, we again factor out a common denominator, without absorbing the exponential factor, and multiply it onto both sides of the equation.
This transforms Eq.~\eqref{eq:KG_r} into the schematic form
\begin{equation}\label{eq:KG_z_01}
e^{- \frac{2 \mu r_+}{1+z}} \left( \sum_{i, j, k, l} \tilde{d}_{i,j,k,l} z^i \chi^j \partial_{z}^k \partial_{\chi}^l \varphi \right) = \sum_{i,j} \tilde{q}_{i,j} z^i \chi^j, 
\end{equation}
where the coefficients $\tilde{d}_{i,j,k,l}$ depend on the original coefficients $d_{i,j,k,l}$, and thus on $M$, $a$, and $\mu$, while $\tilde{q}_{i,j}$ depend only on $q_{i,j}$ and thus only on $M$ and $a$.
As the explicit forms of these coefficients are lengthy and uninformative for understanding the numerical method, we do not present them here.
We emphasize once more that retaining the exponential factor $e^{- \frac{2 \mu r_+}{1+z}}$ explicitly on the derivative side of the equation is a crucial element of our spectral scheme, as it isolates the rapidly varying radial behavior analytically and ensures numerical stability in the subsequent spectral implementation.

We are now ready to perform a spectral expansion to convert Eq.~\eqref{eq:KG_z_01} into a system of algebraic equations.
Formally, an exact representation of $\varphi$ would require an infinite number of spectral basis functions.
In practice, however, one works with a truncated expansion using a finite number of basis functions, which is sufficient to achieve the desired numerical accuracy.
Accordingly, we truncate the spectral expansion to $N+1$ basis functions in both the $z$ and $\chi$ directions. \footnote{
In principle, the truncation orders in the $z$ and $\chi$ directions can be chosen independently.
To maximize the symmetry of the computation procedure and for simplicity, we take them to be equal in this work.}
Because the scalar field $\vartheta$ is even under the parity transformation ($\chi\to-\chi$) in sGB gravity and odd under the same transformation in dCS gravity, $\varphi$ inherits the same parity properties.
We therefore consider the following spectral expansions:
\begin{equation}\label{eq:spectral_expansion_scalar}
\begin{split}
\varphi_{q=1}(z, \chi) & = \sum_{n=0}^{N} \sum_{\ell=0}^{N} \varphi^{(1)}_{n, \ell} T_n (z) P_{2 \ell} (\chi), \\
\varphi_{q=2}(z, \chi) & = \sum_{n=0}^{N} \sum_{\ell=0}^{N} \varphi^{(2)}_{n, \ell} T_n (z) P_{2 \ell+1} (\chi). 
\end{split}
\end{equation}
In what follows, we focus on sGB gravity ($q=1$) to illustrate the procedure, as the implementation for dCS gravity is entirely analogous.
For notational simplicity, we therefore drop the field index and superscripts, with the understanding that we are working in the sGB case. 
Substituting Eq. (\ref{eq:spectral_expansion_scalar}) to (\ref{eq:KG_z_01}), we have 
\begin{equation}\label{eq:KG_z_02}
\begin{split}
& e^{- \frac{2 \mu r_+}{1+z}} \left[ \sum_{i, j, k, l} \tilde{d}_{i,j,k,l} z^i \chi^j \partial_{z}^k \partial_{\chi}^l \left(\sum_{n=0}^{N} \sum_{\ell=0}^{N} \varphi_{n, \ell} T_n (z) P_{2 \ell} (\chi) \right) \right] \\
& = \sum_{i,j} \tilde{q}_{i,j} z^i \chi^j. 
\end{split}
\end{equation}

We can now perform the spectral projection of the Klein–Gordon equation after substituting the spectral expansion of $\vartheta$.
For sGB gravity, the source term satisfies $\mathscr{Q}=\mathscr{G}$ and is even.
Consequently, it suffices to project Eq.~\eqref{eq:KG_z_02} using the even Legendre polynomials in $\chi$.
The right-hand side of Eq.~\eqref{eq:KG_z_02} can be expanded as
\begin{equation}
\begin{split}
& \sum_{i,j} \tilde{q}_{i,j} z^i \chi^j = \sum_{n'=0}^N \sum_{\ell'=0}^N \mathfrak{q}_{n', \ell'} T_{n'} (z) P_{2 \ell'} (\chi), 
\end{split}
\end{equation}
where 
\begin{equation}
\mathfrak{q}_{n', \ell'} = \sum_{i,j} \tilde{q}_{i,j} \int_{-1}^{+1} dz \frac{z^i T_{n'} (z)}{\sqrt{1-z^2}} \int_{-1}^{+1} d \chi \chi^j P_{2 \ell'} (\chi). 
\end{equation}
Similarly, the left-hand side of Eq.~\eqref{eq:KG_z_02} can be expanded as
\begin{equation}
\begin{split}
& e^{- \frac{2 \mu r_+}{1+z}} \left[ \sum_{i, j, k, l} \tilde{d}_{i,j,k,l} z^i \chi^j \partial_{z}^k \partial_{\chi}^l \left(\sum_{n=0}^{N} \sum_{\ell=0}^{N} \varphi_{n, \ell} T_n (z) P_{2 \ell} (\chi) \right) \right] \\
& = \sum_{n'=0}^N \sum_{\ell'=0}^N \left( \sum_{n=0}^N \sum_{\ell=0}^N  \mathfrak{D}_{n'\ell', n\ell} \varphi_{n, \ell} \right) T_{n'} (z) P_{2 \ell'} (\chi) = 0, 
\end{split}
\end{equation}
where
\begin{equation}
\begin{split}
\mathfrak{D}_{n'\ell', n\ell} & = \sum_{i,j,k,l} \tilde{d}_{i,j,k,l} ~ I(i, k, n, n'|2 \mu r_{+}) \\
& \quad \quad \quad \times \int_{-1}^{+1} d \chi \chi^j P^{(l)}_{2 \ell} (\chi) P^{(l)}_{2 \ell'} (\chi)
\end{split}
\end{equation}
Here $I (i, j, k, l| \xi)$ is defined as 
\begin{equation}\label{eq:Integral_func}
I (i, j, k, l| \xi) = \int_{-1}^{+1} dz \frac{z^i T_k^{(j)}(z) T_l (z)}{\sqrt{1-z^2}} \exp \left( -\frac{\xi}{1+z} \right), 
\end{equation}
where $T_k^{(j)}(z)$ stands for the $j$-th derivatives of $T_k(z)$ with respect to $z$. 
This integral is finite for $\xi\geq0$, which is guaranteed in our case since $\xi=2\mu r_{+}>0$.
The convergence of $I(i,j,k,l,|,\xi)$ is a key reason for retaining the exponential factor on the derivative side of the differential operator throughout the construction.
The spectral projection for dCS gravity proceeds in an entirely analogous manner.
The only modification is that all even Legendre polynomials are replaced by odd ones, reflecting the odd parity of the source term $\mathscr{Q}=\mathscr{P}$ under the parity transformation.

By the orthogonality of the Chebyshev and Legendre polynomials, the Klein–Gordon equation is satisfied if
\begin{equation}
\sum_{n=0}^N \sum_{\ell=0}^N  \mathfrak{D}_{n'\ell', n\ell} \varphi_{n, \ell} = \mathfrak{q}_{n', \ell'}. 
\end{equation}
This constitutes an inhomogeneous system of $(N+1)^2$ linear algebraic equations for the spectral coefficients $\varphi{n\ell}$.
For compactness, we introduce the $(N+1)^2\times(N+1)^2$ matrix $\boldsymbol{\mathfrak{D}}$, whose elements are $\mathfrak{D}_{n'\ell',n\ell}$, together with the vectors
\begin{align}
\mathbf{v} &= \Big(v_{0,0}^{i}, v_{1,0}^{i}, \ldots, v_{N,0}^{i}, \ldots, v_{N,1}^{i}, \ldots, v_{N,N}^{i}\Big)^{\rm T}, \nonumber \\
\mathbf{q} &= \Big(\mathfrak{q}_{0,0}^{i}, \mathfrak{q}_{1,0}^{i}, \ldots, \mathfrak{q}_{N,0}^{i}, \ldots, \mathfrak{q}_{N,1}^{i}, \ldots, \mathfrak{q}_{N,N}^{i}\Big)^{\rm T} \nonumber \\
\end{align}
With these definitions, the system can be written compactly as
\begin{equation}
\mathbf{\mathfrak{D}} \mathbf{v} = \mathbf{q}. 
\end{equation}

However, these equations alone are not sufficient to uniquely determine a physical solution for $\vartheta$.
Recall that the scalar field must vanish at spatial infinity, $\vartheta \rightarrow 0$ as $r \rightarrow \infty$, which corresponds to the condition $\vartheta(z=-1)=0$ in the compactified coordinate.
This requirement imposes the following boundary condition at $z=-1$: 
\begin{equation}\label{eq:scalar_constraint}
\sum_{n=0}^{N} (-1)^{n} \varphi_{n \ell} = 0, 
\end{equation}
for all $\ell$. 
These constraints introduce an additional $N+1$ linear equations into the algebraic system.
To incorporate these boundary conditions, we augment the square matrix $\boldsymbol{\mathfrak{D}}$ into a rectangular matrix $\tilde{\boldsymbol{\mathfrak{D}}}$ of dimension $[(N+2)(N+1)] \times (N+1)^2$ and extend the vector $\mathbf{q}$ by appending the corresponding number of zero entries, yielding the augmented vector $\tilde{\mathbf{q}}$.
The resulting system can be written as
\begin{equation}\label{eq:scalar_Eq_aug}
\tilde{\boldsymbol{\mathfrak{D}}} \textbf{v} = \tilde{\textbf{q}}. 
\end{equation}
This system can readily be solved as 
\begin{equation}\label{eq:scalar_Eq_sol}
\textbf{v} = \left( \tilde{\boldsymbol{\mathfrak{D}}}^{\rm T} \tilde{\boldsymbol{\mathfrak{D}}} \right)^{-1} \tilde{\boldsymbol{\mathfrak{D}}}^{\rm T} \tilde{\textbf{q}}, 
\end{equation}
which is equivalent to solving Eq.~(\ref{eq:scalar_Eq_aug}) as a least-square fitting problem. 
In the subsequent section, we verify numerically that the solution obtained from Eq.~(\ref{eq:scalar_Eq_sol}) satisfies the augmented system (\ref{eq:scalar_Eq_aug}) to the desired accuracy.

\subsection{Numerical implementations}

The integral $I(i,j,k,l\mid\xi)$ cannot be evaluated analytically without specifying a branch cut, which would introduce unnecessary ambiguities.
For this reason, we evaluate $I(i,j,k,l\mid\xi)$ numerically throughout this work.
Specifically, for a given value of $\xi$, we compute $I(i,j,k,l\mid\xi)$ using the built-in \textit{Mathematica} function \texttt{NIntegrate}, with a working precision of 700 and accuracy and precision goals set to 350.
We have verified that further increasing the working precision does not lead to any appreciable change in the numerical values, and therefore regard this choice as both accurate and sufficient.

The matrix elements $\mathfrak{D}_{n'\ell',n\ell}$ are then computed from the numerical values of $I(i,j,k,l\mid\xi)$ using a working precision of 300, which is also adopted for constructing the augmented matrix $\tilde{\boldsymbol{\mathfrak{D}}}$.
The inverse of $\tilde{\boldsymbol{\mathfrak{D}}}^{\rm T}\tilde{\boldsymbol{\mathfrak{D}}}$ is computed using the built-in \textit{Mathematica} function \texttt{Inverse} at the same precision.

At a fixed spectral order $N$, the numerical spectral solution generally contains terms that do not vanish in the limit $r \to \infty$.
We refer to these contributions as bare terms, defined by
\begin{equation}
\begin{split}
\varphi_{\rm B} (N) & = \lim_{r \rightarrow + \infty} \varphi (N) \\
& = \text{a polynomial of $\chi$ but not $r^{-1}$}. 
\end{split}
\end{equation}
These terms are unphysical: aside from the exponentially suppressed factor $e^{-\mu r}$, the subleading asymptotic behavior of the scalar field at spatial infinity scales as $r^{-1}$, as derived in Sec.~\ref{sec:massive_scalar_charges}.
We further observe that the coefficients of the bare terms decrease by approximately a factor of $4$–$10$ with each increase in spectral order.
Accordingly, at each spectral order we remove these spurious contributions by subtracting the bare terms,
\begin{equation}
\varphi (N)  - \varphi_{\rm B} (N) \rightarrow \varphi (N), 
\end{equation}
thereby enforcing the correct asymptotic behavior of the scalar field.

\subsection{Numerical results}

In this subsection, we present the numerical results of our spectral construction of massive scalar fields around rotating black holes with various spin parameters.
We begin by showing our convergence and accuracy diagnostics using the representative case $a=0.1$.
Although this corresponds to a relatively small spin, the features observed across the full range of spins considered in this work already emerge.
We then provide a summary of the accuracy of the scalar-field solutions for all spins investigated.
After that, we will show the cross section of the scalar fields. 

\begin{figure*}[htp!]
\centering  
\subfloat{\includegraphics[width=0.47\linewidth]{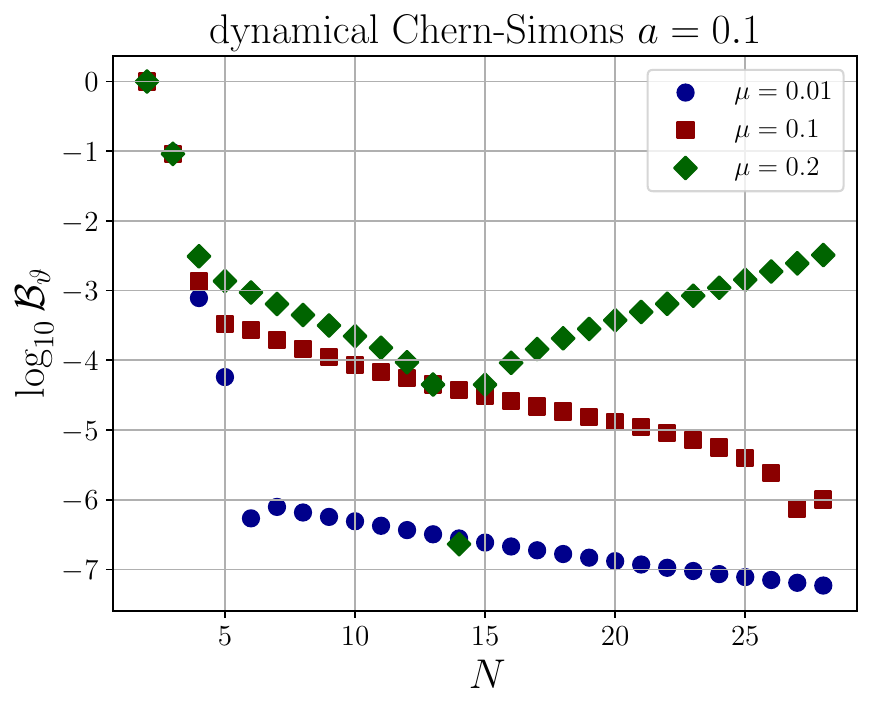}}
\subfloat{\includegraphics[width=0.47\linewidth]{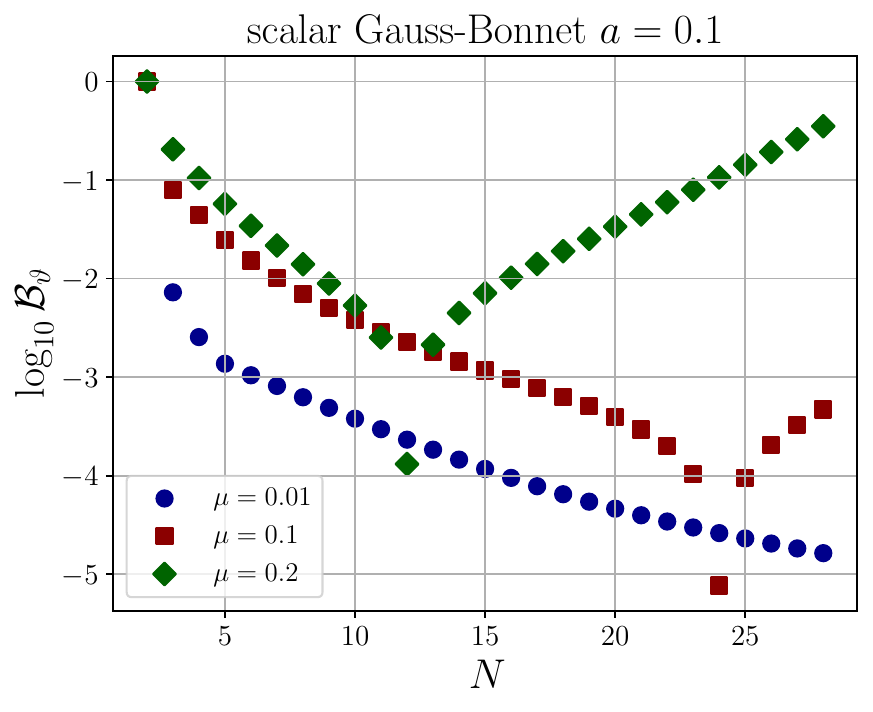}}
\caption{The backward-modulus difference [see Eq.~\eqref{eq:BWD_scalar_field} in the main text for definition] of the massive scalar field of mass $\mu = 0.01$ (dark-blue circles), 0.1 (dark-red squares) and 0.2 (dark-green diamonds) around a rotating black hole of dimensionless spin $a=0.1$ in dynamical Chern-Simons (dCS, left panel) and scalar Gauss-Bonnet (sGB, right panel) gravity as a function of the spectral order $N$. 
}
\label{fig:BMD_scalar_a_010}
\end{figure*}

\subsubsection{Backward modulus difference}

To assess the convergence of the scalar field $\vartheta$ with increasing spectral order $N$, we define the backward modulus difference (BMD) as
\begin{equation}\label{eq:BWD_scalar_field}
    \mathcal{B}_{\vartheta}(N) = \left[ \int_{r_+}^{\infty} \int_{-1}^{+1} \left[ \varphi(N)-\varphi(N-1)\right]^2 \, dr d\chi \right]^{1/2}. 
\end{equation}
The exponential factor $e^{-\mu r}$ is deliberately excluded from this definition, since it deforms $\varphi(N)$ and $\varphi(N-1)$ in the same manner and therefore does not affect their relative difference.
Moreover, because $e^{-\mu r}$ decays faster than any inverse power of $r$ as $r\to\infty$, omitting this factor allows the BMD to expose potential divergences that might otherwise be artificially suppressed.
An additional advantage of this definition is that $\mathcal{B}_{\vartheta}$ can be evaluated analytically term by term, as $\varphi$ is expressed as a power series in $r^{-1}$ and $\chi$, with coefficients obtained numerically from Eq.~\eqref{eq:scalar_Eq_sol}.

Figure~\ref{fig:BMD_scalar_a_010} shows the BMD of $\varphi$ for scalar-field masses $\mu=0.01$ (dark-blue circles), $0.1$ (dark-red squares), and $0.2$ (dark-green diamonds) around a rotating black hole with dimensionless spin $a=0.1$ in dCS (left panel) and sGB gravity (right panel), as a function of spectral order $N$.
Several features observed in Fig.~\ref{fig:BMD_scalar_a_010} also emerge from the results at other spin values.
For both gravity theories, $\mathcal{B}_{\vartheta}$ initially decreases approximately exponentially with $N$ up to a characteristic spectral order, beyond which it continues to decrease at a different exponential rate (for $\mu < 0.1$).
We denote this transition point as the twist spectral order, $N_{\rm (twist)}$, and find that $N_{\rm (twist)}$ decreases as the scalar-field mass $\mu$ increases.
For $\mu\geq0.1$, $\mathcal{B}_{\vartheta}$ also reaches a minimum at a spectral order $N_{\rm (min)}$, and increases again.
As with $N_{\rm (twist)}$, the value of $N_{\rm (min)}$ also shifts to lower spectral order as $\mu$ increases.

The emergence of the characteristic spectral orders $N_{\rm (twist)}$ and $N_{\rm (min)}$ can be understood as a consequence of the radial variation introduced by the exponentially decaying factor $e^{-\mu r}$.
The mass of the scalar field introduces a characteristic length scale $\lambdabar$, which is asymptotically proportional to the Compton wavelength of the field, $\lambdabar \sim \mathcal{O}(\mu^{-1})$.
This scale effectively partitions the black-hole spacetime into two regions with qualitatively different behavior.
The first region corresponds to $r\in(r_+,\lambdabar]$, while the second corresponds to $r\in[\lambdabar,\infty)$.

In terms of the compactified radial coordinate $z$, these two regions map to $z\in[\bar z,1)$ and $z\in(-1,\bar z]$, respectively, where
\begin{equation}
\bar{z} = \frac{2 r_+}{\lambda} -1 \sim \mathcal{O} \left( \mu r_+ -1 \right). 
\end{equation}
Within the inner region $z\in[\bar z,1)$ [equivalently, $r\in(r_+,\lambdabar]$], the scalar field is dominated by the nonminimal coupling.
In this regime, the exponential factor varies slowly, with $e^{-\mu r}=1+\mathcal{O}(\mu)$, and the scalar field remains relatively smooth.
This smoothness explains the faster exponential convergence observed for spectral orders below $N_{\rm (twist)}$.

By contrast, in the outer region $z\in(-1,\bar z]$ [i.e., $r\in[\lambdabar,\infty)$], the scalar field is dominated by the screening of its own mass, as the exponential suppression $e^{-\mu r}$ overwhelms all inverse power-law tails.
In the compactified coordinate, the quantity $1+\bar z\sim\mu r_+$ can be interpreted as the effective ``distance" between the length scale $\lambdabar$ and spatial infinity in $z$-space.
As the spectral order $N$ increases, the characteristic resolution scale of the Chebyshev polynomials decreases.
Once this scale becomes smaller than $1+\bar z$, the spectral basis begins to resolve regions with sharply different characteristics simultaneously.
At this point, increasing $N$ no longer improves the global approximation uniformly, and the BMD reaches a minimum at $N_{\rm (min)}$.

Since the characteristic resolution scale of Chebyshev polynomials scales approximately as $N^{-1}$, one can estimate $N_{\rm (min)}$ by equating this scale with $1+\bar z$, yielding
\begin{equation}
\begin{split}
N_{\rm (min)} \sim \mathcal{O} \left( \frac{1}{\mu r_+}\right). 
\end{split}
\end{equation}
This scaling is consistent with the numerical results shown in Fig.~\ref{fig:Err_scalar_a_010}.
For example, focusing on the dCS case, we find $N_{\rm (twist)} \sim 27$ for $\mu=0.1$ and $N_{\rm (twist)}\sim14$ for $\mu=0.2$, in agreement with the inverse-$\mu$ scaling predicted by the above estimate.

Finally, we note that the spectral method is applied to the rescaled field $\varphi$, from which the exponential factor $e^{-\mu r}$ has been factored out, and which is therefore smoother than $\vartheta$ itself.
Nevertheless, as seen explicitly in Eqs.~(\ref{eq:KG_r}), (\ref{eq:KG_z_01}), and (\ref{eq:KG_z_02}), the exponential factor remains attached to the differential operator.
As a result, the presence of $e^{-\mu r}$ continues to influence the numerical construction of $\vartheta$, ultimately limiting the achievable spectral convergence at large $\mu$.

\subsubsection{Absolute error}

To quantify how well the scalar field $\vartheta$ constructed using the spectral scheme satisfies the Klein–Gordon equation, $E_{\vartheta}=0$, we define the following absolute error, based on the definition in Refs.~\cite{Lam:2025fzi,Lam:2025elw}:
\begin{equation}\label{eq:Err_scalar}
\mathcal{E}_{\vartheta} \propto \left[ \int_{r_+}^{\infty} \int_{-1}^{+1} E_{\vartheta}^{2} \left[-g^{(0)}\right]^{6} \, \frac{dr}{r^{26}} d\chi \right]^{\frac{1}{2}}. 
\end{equation}
If $\vartheta$ were an exact solution of the Klein–Gordon equation, then $\mathcal{E}_{\vartheta}$ would vanish identically.
Since the numerical solution obtained via the spectral method is approximate, $\mathcal{E}_{\vartheta}$ is nonzero, and its magnitude provides a measure of the extent to which $\vartheta$ satisfies $E_{\vartheta}=0$ throughout spacetime.

In Eq.~\eqref{eq:Err_scalar}, the factor $\left(-g^{(0)}\right)^{6}/r^{26}$ serves as a regularization weight that ensures the convergence of the integral.
Before motivating this specific choice, we first justify the legitimacy of introducing such a factor.
There is no unique or unambiguous way to define a residual norm of an equation, as multiplying a differential equation by any nonvanishing function yields an equivalent equation.
Thus, it is the relative variation of $\mathcal{E}_{\vartheta}$ with spectral order, rather than its absolute value, that carries physical and numerical significance.
By this token, we fix the overall proportionality constant such that $\mathcal{E}_{\vartheta}(N=1)=1$ for different $\mu$.
We have also explored alternative definitions of the residual norm, and find that all reasonable choices exhibit qualitatively similar convergence behavior as a function of spectral order.
Our conclusions are therefore insensitive to the particular normalization adopted in Eq.~\eqref{eq:Err_scalar}.

A key advantage of introducing the regularization factor is that the integrand in Eq.~\eqref{eq:Err_scalar} can be expanded as a series of the following form
\begin{equation}
\begin{split}
& \frac{\left[-g^{(0)}\right]^{6}}{r^{26}}E_{\vartheta}^{2} = \sum_{j=2} \sum_{k=0} \sum_{q=0}^{2} \delta_{j,k,q} \frac{\exp(-q \mu r)}{r^{j}}\chi^k, 
\end{split}
\end{equation}
where the coefficients $\delta_{j,k,q}$ are constants that can be computed algebraically from the spectral coefficients $v_{n\ell}$.
Each term in this expansion can be integrated analytically using the identity, 
\begin{equation}\label{eq:math_trick}
\begin{split}
\int_{r_+}^{+ \infty} dr \frac{\exp(-q \mu r)}{r^{j}} & = \frac{1}{r_{+}^{j-1}} \text{Ei}_{j} (q \mu r_+), 
\end{split}
\end{equation}
where $\text{Ei}_{j}$ is the exponential integral, defined as 
\begin{equation}
\text{Ei}_{j} (z) = \int_{1}^{+\infty} dx \frac{e^{-zx}}{x^j}. 
\end{equation}
This analytic treatment eliminates the numerical errors that would otherwise arise from direct numerical integration of the residual norm.

\begin{figure*}[htp!]
\centering  
\subfloat{\includegraphics[width=0.47\linewidth]{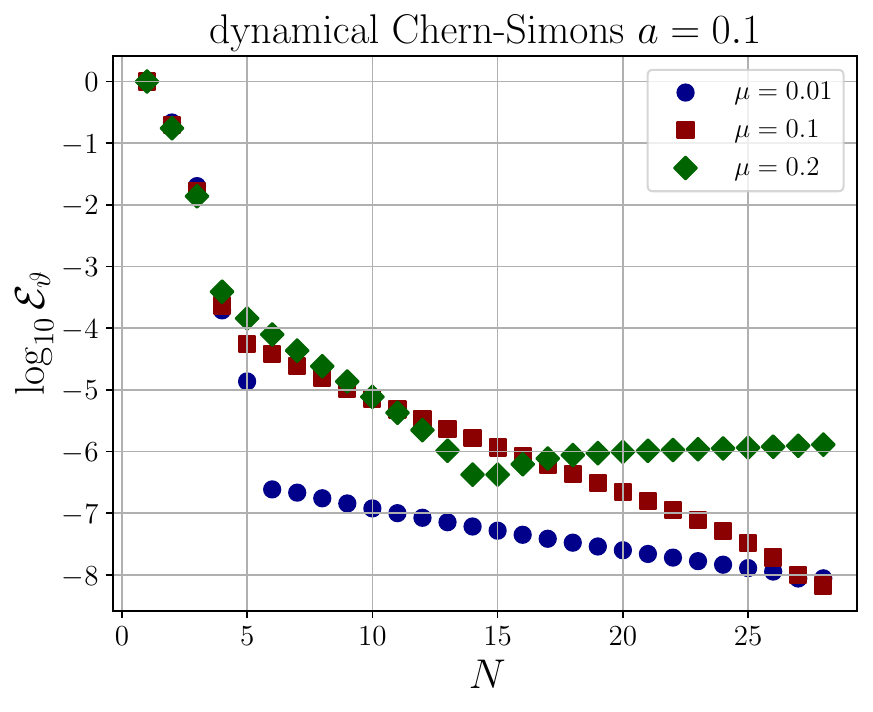}}
\subfloat{\includegraphics[width=0.47\linewidth]{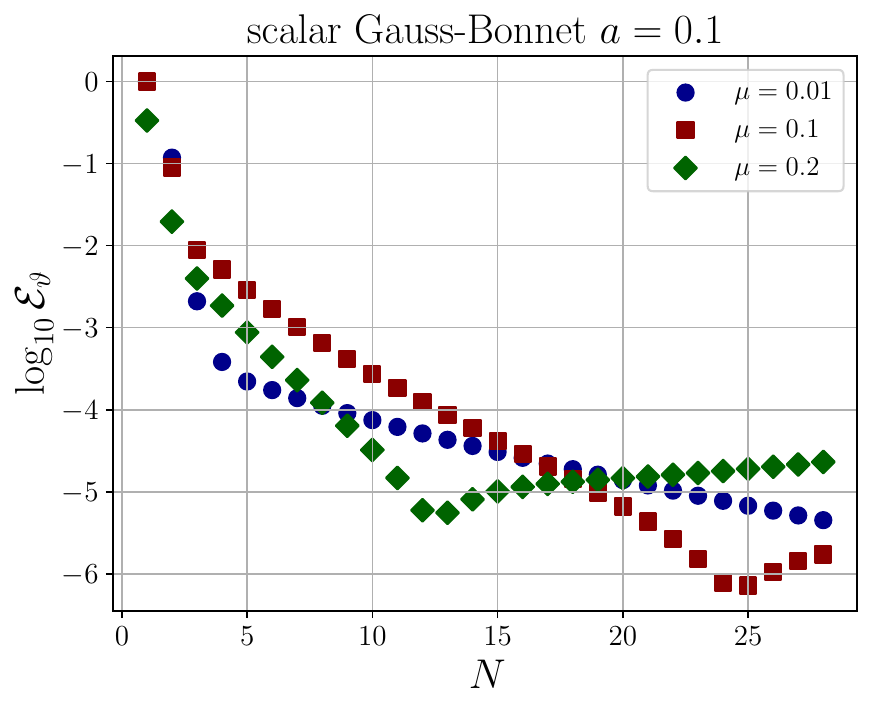}}
\caption{The error [see Eq.~(\ref{eq:Err_scalar}) in the main text for definition] of the massive scalar field of mass $\mu = 0.01$ (circles in dark blue), 0.1 (squares in dark red) and 0.2 (diamonds in dark green) around a rotating black hole of dimensionless spin $a=0.1$ in dCS (left panel) and sGB (right panel) gravity as a function of the spectral order $N$. 
As there is no a universally unambiguous definition of the residual of an equation, we have normalized the residual at $N = 1$ to be unity for different $\mu$. 
We observe that, for all $\mu$ and both gravity theories, the error first exponentially decrease at a rate as $N$ increases to a spectral order, and then continues to decrease exponentially, but with a smaller rate. 
For $\mu=0.1$ and 0.2, the error could even reach a minimum for $N \leq 30$.  
These patterns are related to the radial variation of the scalar field introduced by the exponential factor $e^{-\mu r}$. 
}
\label{fig:Err_scalar_a_010}
\end{figure*}

Figure~\ref{fig:Err_scalar_a_010} shows the error $\mathcal{E}_{\vartheta}$ as a function of spectral order $N$ for a rotating black hole with $a=0.1$, for scalar-field masses $\mu = 0.01$ (blue circles), 0.1 (red squares) and 0.2 (green diamonds) in dCS (left panel) and sGB gravity (right panel).
We find that $\mathcal{E}_{\vartheta}$ exhibits convergence behavior qualitatively similar to that of the backward modulus difference $\mathcal{B}_{\vartheta}$.
In particular, $\mathcal{E}_{\vartheta}$ initially decreases exponentially with spectral order up to a characteristic value of $N$, beyond which it continues to decrease exponentially but at a different rate.
As $\mu$ increases, $\mathcal{E}_{\vartheta}$ also reaches a minimum at a finite spectral order.
Importantly, the spectral orders at which $\mathcal{E}_{\vartheta}$ and $\mathcal{B}_{\vartheta}$ change their convergence rates and attain their minima at similar spectral orders, as expected for consistent measures of error.

To minimize the propagation of numerical error in subsequent calculations involving the scalar field, we select the solution that best satisfies $E_{\vartheta}=0$.
For a given scalar-field mass $\mu$ and dimensionless $a$, this is achieved by choosing the spectral order at which the error $\mathcal{E}_{\vartheta}$ attains its minimum.
We denote this optimal spectral order by $N_{\rm opt}$, defined as
\begin{equation}\label{eq:opt_spec_order_phi} 
N_{\rm opt} = \text{arg}\min_{N} \mathcal{E}_{\vartheta},
\end{equation}
and refer to the corresponding value $\mathcal{E}_{\vartheta}(N_{\rm opt})$ as the least error.
The scalar field $\vartheta(N_{\rm opt})$ is then adopted for all downstream calculations in order to minimize errors.

\subsubsection{Results for $a \leq 0.8$}

\begin{figure*}[htp!]
\centering  
\subfloat{\includegraphics[width=0.47\linewidth]{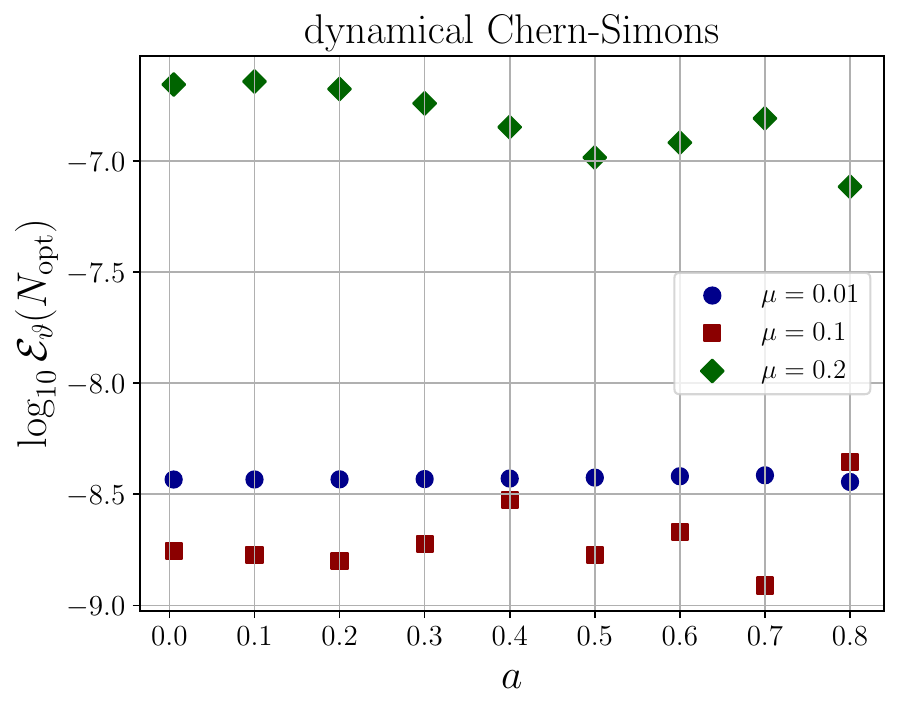}}
\subfloat{\includegraphics[width=0.47\linewidth]{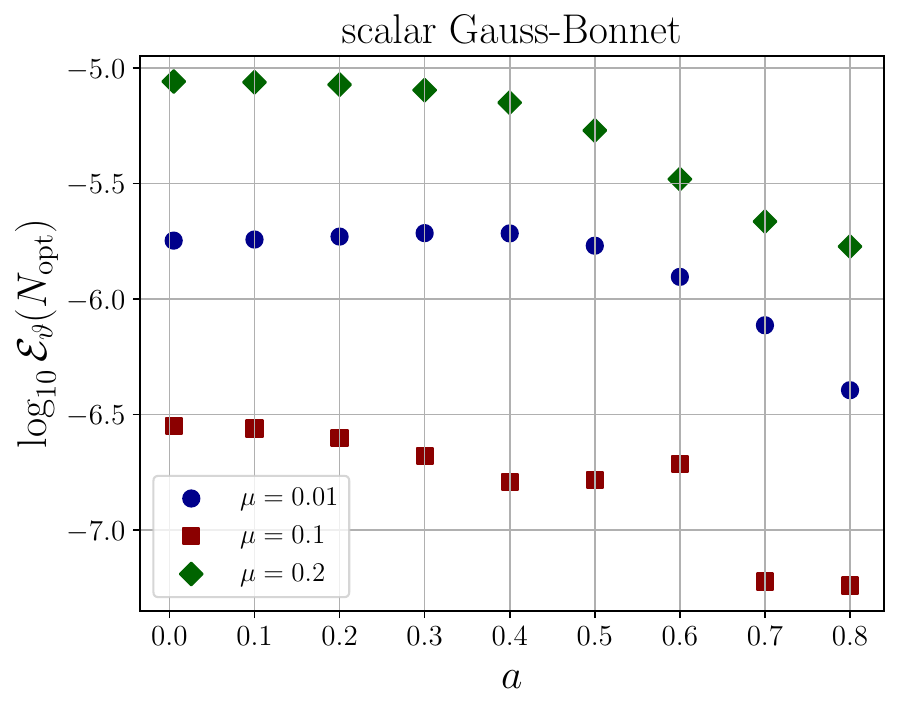}}
\caption{The least error [see the text around Eq.~\eqref{eq:opt_spec_order_phi} for definition] of the massive scalar field of mass $\mu = 0.01$ (circles in dark blue), 0.1 (squares in dark red) and 0.2 (diamonds in dark green) around a rotating black hole in dCS (left panel) and sGB (right panel) gravity as a function of the dimensionless spin $a$ 
We observe that, relatively, the least error show less variations over $a$, but increase significantly as $\mu$ increases. 
This change can be explained from the rapid changes introduced to the scalar field profile by the exponential factor $e^{-\mu r}$ as $\mu$ increases. 
}
\label{fig:Min_err_scalar_a}
\end{figure*}

Figure~\ref{fig:Min_err_scalar_a} shows the least error, $\mathcal{E}_{\vartheta}(N_{\rm opt})$, as a function of the dimensionless spin $a$ for $\mu = 0.01$, $0.1$, and $0.2$ in dCS (left panel) and sGB (right panel) gravity.
We find that the least error exhibits no significant dependence on the spin parameter $a$, but increases noticeably as the scalar-field mass is raised from $\mu = 0.1$ to $\mu = 0.2$.
This behavior is consistent with the discussion above regarding the intrinsic limitation on achieving $E_{\vartheta}=0$, which arises from the presence of the exponentially decaying factor $e^{-\mu r}$ in the differential operator governing the equation for $\vartheta$.
As $\mu$ increases, the influence of the $e^{-\mu r}$ term becomes more pronounced, introducing sharper radial features in $\vartheta$ that are more difficult to resolve spectrally, thereby reducing the overall accuracy of the spectral method at $\mu = 0.2$.

\begin{figure*}[htp!]
\centering  
\subfloat{\includegraphics[width=0.31\linewidth]{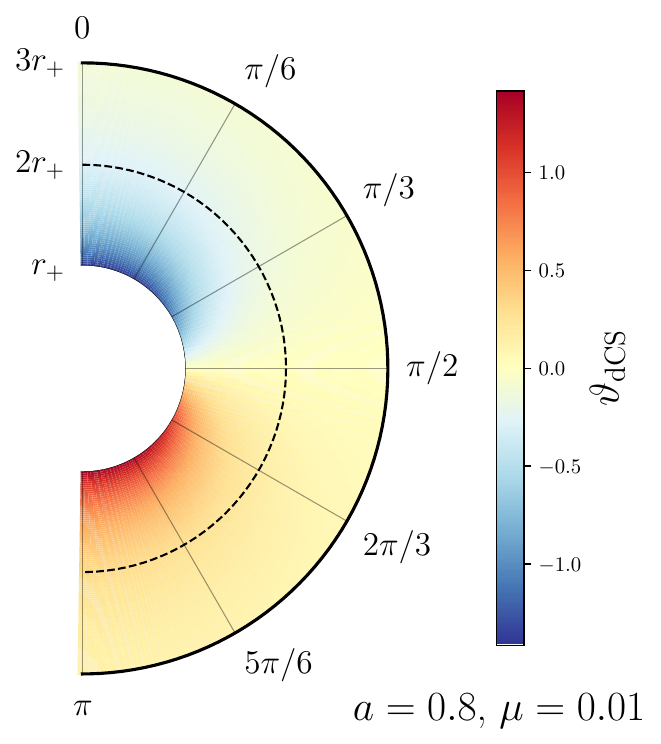}}
\subfloat{\includegraphics[width=0.31\linewidth]{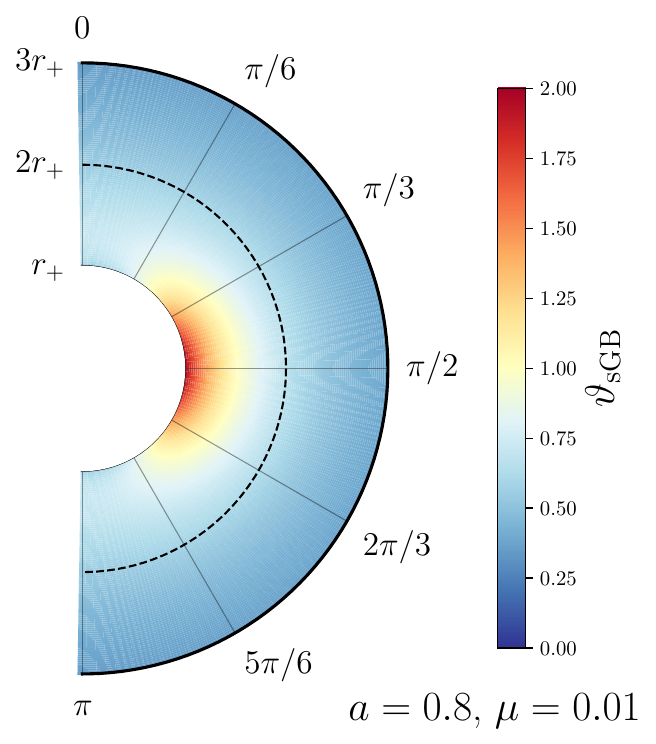}}
\qquad
\subfloat{\includegraphics[width=0.31\linewidth]{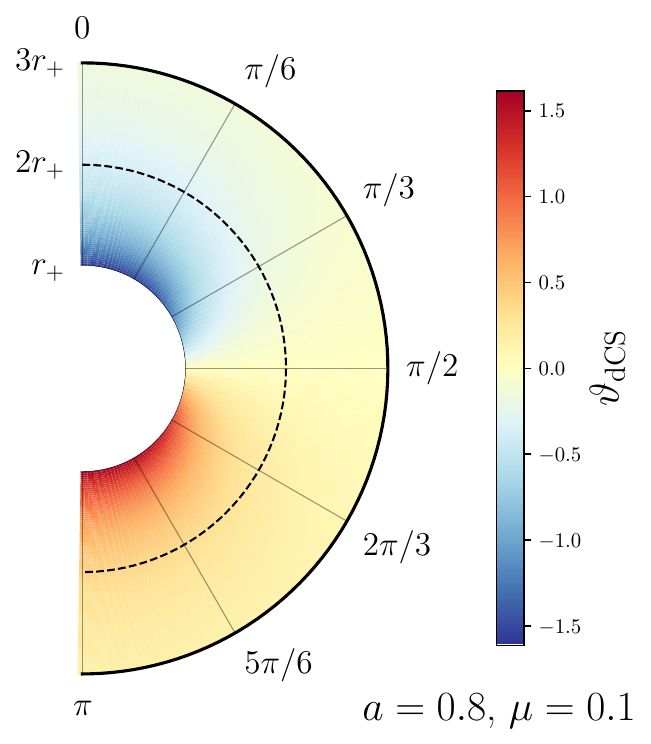}}
\subfloat{\includegraphics[width=0.31\linewidth]{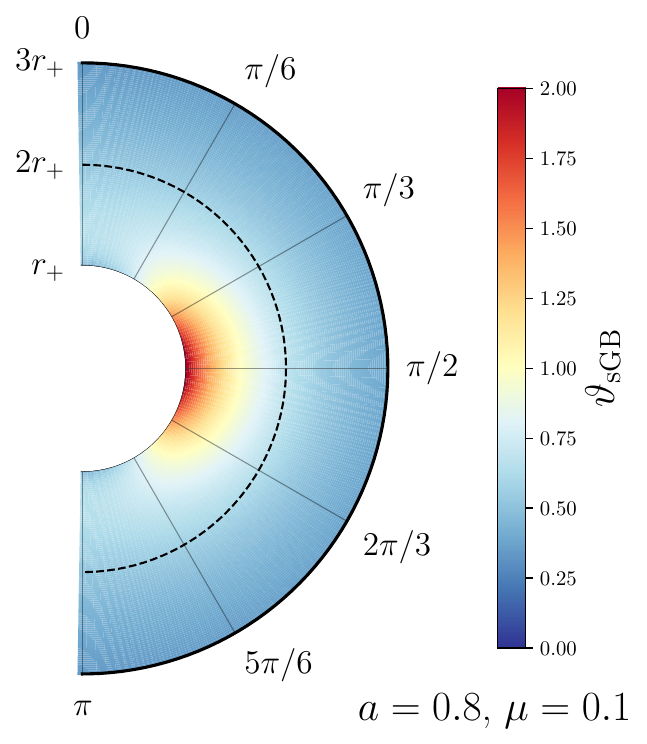}}
\qquad
\subfloat{\includegraphics[width=0.31\linewidth]{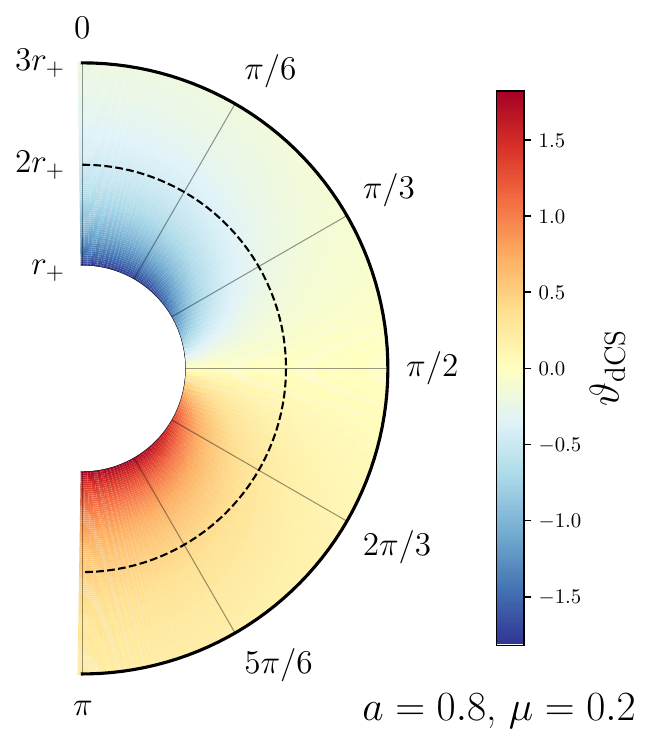}}
\subfloat{\includegraphics[width=0.31\linewidth]{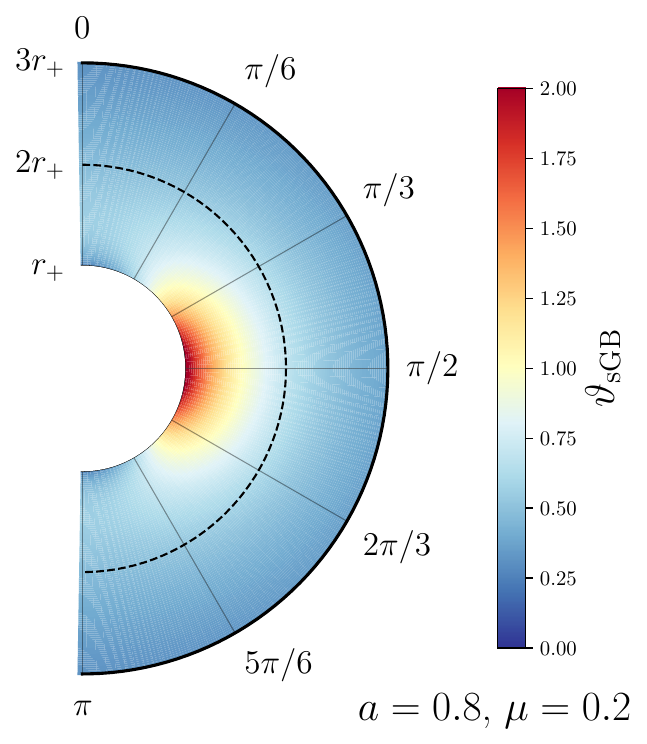}}
\caption{Meridional cross sections of the massive scalar field around a rotating black hole with dimensionless spin $a = 0.8$ in dCS (left panels) and sGB (right panels) gravity. 
The top, middle, and bottom panels correspond to scalar-field masses $\mu = 0.01$, $\mu = 0.1$, and $\mu = 0.2$, respectively. 
All solutions are computed at their respective optimal spectral resolutions. 
In dCS gravity, the scalar field exhibits a dipolar structure and is antisymmetric about the equatorial plane ($\chi = 0$, or $\theta = \pi/2$), which acts as a nodal plane where the field vanishes at all radii. 
In contrast, the scalar field in sGB gravity displays a quadrupolar structure and is symmetric about the equator. 
In both theories, the radial profile decays exponentially with increasing radius due to the factor $e^{-\mu r}$. 
The overall multipolar structure of the scalar field is not significantly altered by the presence of a finite mass, consistent with previous massless results \cite{Lam:2025fzi}.
}
\label{fig:Scalar_meridian}
\end{figure*}

Figure~\ref{fig:Scalar_meridian} shows meridional cross section plots of the massive scalar field around a rotating black hole with dimensionless spin $a = 0.8$ in dCS (left panels) and sGB (right panels) gravity.
The top, middle, and bottom panels correspond to scalar-field masses $\mu = 0.01$, $\mu = 0.1$, and $\mu = 0.2$, respectively.
All scalar fields are computed using their respective optimal spectral resolutions.
We present results at this spin value to clearly illustrate the multipolar structure of the massive scalar fields.
In dCS gravity, the scalar field exhibits a dominant dipolar structure, whereas in sGB gravity it displays a clear quadrupolar structure.
In particular, for dCS gravity the equator ($\chi = 0$, or equivalently $\theta = \pi/2$) is a nodal plane, where the scalar field vanishes at all radii.
The scalar field is also antisymmetric about this plane, reflecting the antisymmetry of the Pontryagin density $\mathscr{P}$.
By contrast, no such nodal plane exists in sGB gravity, and the scalar field is symmetric about the equator, consistent with the fact that the Gauss–Bonnet invariant $\mathscr{G}$ is symmetric under the transformation $\chi \rightarrow -\chi$.
In both theories, the scalar field decays more rapidly with increasing radius as the field mass increases, due to the exponentially decaying factor $e^{-\mu r}$.
However, the decay is not readily apparent, and the scalar-field profiles shown exhibit only a weak dependence on $\mu$. 
This behavior can be understood from the fact that, within the near-hole region (i.e., for $r \lesssim \mu^{-1}$), the scalar field is primarily sourced by the curvature invariant and therefore closely resembles the corresponding massless solution. 
Increasing $\mu$ mainly affects the large-radius behavior by introducing stronger exponential (Yukawa) suppression.
Finally, from Fig.~\ref{fig:Scalar_meridian}, and in comparison with the massless results reported in Ref.~\cite{Lam:2025fzi}, we conclude that the overall geometry of the scalar field, namely, the relative composition of its multipolar structure, is not significantly modified by the presence of a finite scalar-field mass.

%%%%%%%%%%%%%%%%%%%%%%%%%%%%%%%%%%%%%%%%%%%%%%%%%%%%%%%%%%%%%%%%%%%%%%%%%%%%%%%%%%%%%%%%%%%%%%%%%%%%%%%%%%%%%%%%%%%%%%%%%%%%
\section{Constructing the modifications to the background spacetime}\label{sec:Metric}

In this section, we develop spectral methods to solve the modified Einstein equations for the modifications to black-hole spacetimes due to the presence of massive scalar charges. 
We begin by outlining the analytical framework underlying our spectral approach to solving the modified Einstein equation.
We then describe the details of the numerical implementation.
Finally, we present and analyze the numerical results concerning the spacetime modifications.

\subsection{Spectral methods}

We now solve the modified field equations $E_{\beta}{}^{\nu} = 0$. 
Schematically, these equations can be expressed as \cite{Lam:2025fzi}
\begin{equation}\label{eq:EE1Schematic}
\sum_{i = 1}^{4} \ (\mathscr{D}_{i})_{\beta}{}^{\nu} H_i = - [\mathscr{A}_{\beta}{}^{\nu}]^{(0)} + [\bar{T}_{\beta}{}^{\nu}]^{(0)},  
\end{equation}
where $(\mathscr{D}_{i})_{\beta}{}^{\nu}$ is a linear differential operator. 
This system consists of coupled two-dimensional partial differential equations in the variables $r$ and $\chi$.
By construction, the metric ansatz in Eq.~\eqref{eq:metric} already satisfies the $(\mu,\nu)=(t,r)$, $(t,\chi)$, $(r,\phi)$, and $(\chi,\phi)$ components of the modified field equations.
We therefore solve the remaining six components simultaneously using spectral methods.

Before extracting a common denominator and factorizing the equations, we first change the radial coordinate from $r$ to the compactified coordinate $z$.
Since $\mathscr{A}_{\beta}{}^{\nu}$ and $\bar{T}_{\beta}{}^{\nu}$ depend linearly and quadratically on the scalar field $\vartheta$ and its derivatives, respectively, their dependence in the $z$ coordinate (at fixed $\chi$) takes the form
\begin{equation}
\begin{split}
\mathscr{A}_{\beta}{}^{\nu} & \propto e^{-\frac{2 \mu r_+}{1+z}}, \\
\bar{T}_{\beta}{}^{\nu} & \propto e^{-\frac{4 \mu r_+}{1+z}}. 
\end{split}
\end{equation}
By construction, the rescaled field $\varphi$ is expressed as a polynomial in $z$ and $\chi$.
As a result, the quantities $R_{\beta}{}^{\nu}$, $e^{\frac{2\mu r_+}{1+z}}\mathscr{A}_{\beta}{}^{\nu}$, and $e^{\frac{4\mu r+}{1+z}}\bar{T}_{\beta}{}^{\nu}$ contain only rational functions of $z$ and $\chi$.
Thus, we extract the least common denominator shared by these three expressions and multiply it through both sides of the equations.
Through this procedure, the modified field equations can be cast into the form, 
\begin{equation}\label{eq:EE1z}
\begin{split}
& \sum_{i=1}^{4} \sum_{\delta,\sigma} \sum_{\alpha,\beta=0}^{2} \mathcal{G}_{i,\delta,\sigma,\alpha,\beta}^{j} z^{\delta} \chi^{\sigma} \partial_{z}^{\alpha} \partial_{\chi}^{\beta} H_i(z,\chi) \\
& = e^{-\frac{2 \mu r_+}{1+z}} \sum_{\delta,\sigma} \mathcal{A}_{\delta,\sigma}^{j} z^{\delta} \chi^{\sigma} + e^{-\frac{4 \mu r_+}{1+z}} \sum_{\delta,\sigma} \mathcal{T}_{\delta,\sigma}^{j} z^{\delta} \chi^{\sigma}. 
\end{split}
\end{equation}
where the index $j=1,\dots,6$ labels the independent components of the field equations.
The coefficients $\mathcal{G}^{j}_{i,\delta,\sigma,\alpha,\beta}$ depend only on the background parameters $M$ and $a$, while $\mathcal{A}^{j}_{\delta,\sigma}$ and $\mathcal{T}^{j}_{\delta,\sigma}$ depend on the scalar-field spectral coefficients $c_{n\ell}$, as well as on $M$, $a$, and $\mu$.
Here $\mathcal{A}^{j}_{\delta,\sigma}$ originate from $\mathscr{A}_{\beta}{}^{\nu}$, whereas $\mathcal{T}^{j}_{\delta,\sigma}$ arise from $\bar{T}_{\beta}{}^{\nu}$.
A key feature of our spectral formulation is that the exponential factors $e^{-\frac{2\mu r_+}{1+z}}$ and $e^{-\frac{4\mu r_+}{1+z}}$ are retained in the source terms of the differential equations.
This procedure plays an essential role in achieving stable and accurate numerical solutions.

We now perform a spectral expansion of the metric modifications $H_i$.
As in the case of the scalar field $\varphi$, an exact spectral representation of $H_i(z,\chi)$ would formally require an infinite number of basis functions.
In practice, however, and following the same truncation strategy adopted for $\varphi$, we approximate $H_i$ using a finite set of $N+1$ spectral bases in both the $z$ and $\chi$ directions,
\begin{equation}\label{eq:HAnsatz}
H_i(z,\chi) = \sum_{n=0}^{N} \sum_{\ell=0}^{N} v_{n\ell}^{i} T_{n}(z) P_{\ell}(\chi). 
\end{equation}
We note that, unlike in Refs.~\cite{Lam:2025elw,Lam:2025fzi}, where only even Legendre polynomials were involved in the expansion of $H_i$, we retain both even and odd Legendre polynomials, despite the fact that the functions $H_i(z,\chi)$ themselves are even in $\chi$.
This procedure is necessitated by the inclusion of the $(\beta,\nu)=(r,\chi)$ component of the modified field equations, which is odd under $\chi\rightarrow-\chi$.
A consistent spectral representation of this equation therefore requires the inclusion of odd Legendre polynomials.
Substituting the spectral expansion \eqref{eq:HAnsatz} into Eq.~\eqref{eq:EE1z}, projecting the resulting equations onto the product basis $T_n(z)P_{\ell}(\chi)$, and following the procedure described in Ref.~\cite{Lam:2025fzi}, we obtain a coupled system of linear algebraic equations for the coefficients $v^{i}_{n\ell}$., 
\begin{equation}
\begin{split}
\sum_{n=0}^{\mathcal{N}_z} \sum_{\mathrm{even} \, \ell}^{\mathcal{N}_{\chi}} \sum_{i=1}^{4} [\mathbb{D}_{j n' \ell', i n \ell}] v_{n\ell}^{i} = s_{n'\ell'}^{j}. 
\end{split}
\end{equation} 
Here, $\mathbb{D}_{j n' \ell', i n \ell}$ is defined by Eq. (46) in \cite{Lam:2025fzi}, and $s_{n'\ell'}^{j}$ can be evaluated in terms of $I(i, j, k, l|\zeta)$ defined by Eq.~(\ref{eq:Integral_func}), 
\begin{equation}\label{eq:source_compt}
\begin{split}
s_{n'\ell'}^{j} = \sum_{\delta, \sigma} \Big[ & \mathcal{A}_{\delta,\sigma}^{j} I(\delta,0,0,n'|2 \mu r_+) \\
& + \mathcal{T}_{\delta,\sigma}^{j} I(\delta,0,0,n'|4 \mu r_+) \Big] \\
& \times \int_{-1}^{+1} d \chi \chi^{\sigma} P_{2 \ell}(\chi), 
\end{split}
\end{equation}
where $I(\delta,0,0,n'|2 \mu r_+)$ and $I(\delta,0,0,n'|4 \mu r_+)$ are defined by Eq.~\eqref{eq:Integral_func}. 
This system of the algebraic equations can be compactly denoted as 
\begin{equation}\label{eq:MatrixEq}
\mathbb{D} \mathbf{v} = \mathbf{s}, 
\end{equation}
where 
\begin{align}
\mathbf{v} &= \Big(v_{0,0}^{i}, v_{1,0}^{i}, \ldots, v_{N,0}^{i}, \ldots, v_{N,1}^{i}, \ldots, v_{N,N}^{i}\Big)^{\rm T}, \nonumber \\
\mathbf{s} &= \Big(s_{0,0}^{i}, s_{1,0}^{i}, \ldots, s_{N,0}^{i}, \ldots, s_{N,1}^{i}, \ldots, s_{N,N}^{i}\Big)^{\rm T}, \nonumber \\
\end{align}
and $\mathbb{D}$ is a $[4(N+1)]^2 \times [4(N+1)]^2$ matrix, and $\textbf{v}$ and $\textbf{s}$ are both a $[4(N+1)]^2$ vector. 

As in the scalar-field case, Eq.~\eqref{eq:MatrixEq} alone is insufficient to uniquely determine the physical metric modifications.
The system must be supplemented by appropriate boundary conditions, Eq.~\eqref{eq:BoundaryConditions}, which in the compactified $z$ coordinate take the form \cite{Lam:2025elw,Lam:2025fzi}
\begin{equation}\label{eq:BoundaryConditions}
\begin{split}
    \sum_{n=0}^{N} (-1)^n v_{n\ell}^{1} &= 0, \\
    \sum_{n=0}^{N} (-1)^n v_{n\ell}^{2} &= r_+ \sum_{n=0}^{N} (-1)^{n+1} n^2 v_{n\ell}^{3}, \\
    \sum_{n=0}^{N} (-1)^n v_{n\ell}^{3} &= -2r_+ \sum_{n=0}^{N} (-1)^{n+1} n^2 v_{n\ell}^{3}, \\
    \sum_{n=0}^{N} (-1)^n v_{n\ell}^{4} &= -2r_+ \sum_{n=0}^{N} (-1)^{n+1} n^2 v_{n\ell}^{3}. 
\end{split}
\end{equation}
These add $4(N+1)$ linear algebraic equations to Eq.~(\ref{eq:MatrixEq}). 

Unlike the approach adopted in Ref.~\cite{Lam:2025fzi}, where the boundary conditions are first used to eliminate a subset of the spectral coefficients before solving the reduced system, we instead incorporate these constraints directly into the linear system.
Specifically, we augment the coefficient matrix $\mathbb{D}$ to form an enlarged matrix $\tilde{\mathbb{D}}$, and extend the source vector $\mathbf{s}$ to $\tilde{\mathbf{s}}$ by appending zeros at the corresponding entries to enforce the boundary conditions.
The resulting augmented system can be written compactly as
\begin{equation}\label{eq:MatrixEq2}
\tilde{\mathbb{D}} \mathbf{v} = \tilde{\mathbf{s}}. 
\end{equation}
We then solve this overdetermined system by 
\begin{equation}
\mathbf{v} = \left( \tilde{\mathbb{D}}^{\rm T} \tilde{\mathbb{D}} \right)^{-1} \tilde{\mathbb{D}}^{\rm T} \tilde{\textbf{s}}. 
\end{equation}
By substituting the elements of $\mathbf{v}$ back to the spectral expansion of $H_i(z, \chi)$ and changing the variable from $z$ to $r$, we can construct the spacetime of rotating black holes surrounded by massive scalar charges, and $H_i(r, \chi)$ in axi-dilaton gravity could be obtained by summing these functions in dCS and sGB gravity. 

We conclude this subsection with a brief remark on the structure of $\tilde{\mathbf{s}}$.
From Eq.~\eqref{eq:source_compt}, we see that each component of $\tilde{\mathbf{s}}$ comprises two contributions, both expressed in terms of the integral function $I$ defined in Eq.~\eqref{eq:Integral_func}.
As a result, $\tilde{\mathbf{s}}$ encodes information obtained by convolving the source terms of the modified Einstein equations over the entire exterior region of the black hole spacetime, from the event horizon to spatial infinity.
In this sense, the appearance of the integral function $I$ in $\tilde{\mathbf{s}}$ plays a role analogous to that of the integrals encountered in solving the Tolman-Oppenheimer-Volkoff equation for stellar spacetimes~\cite{Oppenheimer:1939ne}.

\subsection{Numerical implementations}

The numerical implementations for computing $H_i(r, \chi)$ are largely similar to those for constructing the scalar fields.
The integral function $I$ is computed using the built-in \textit{Mathematica} function \texttt{NIntegrate}, with a working precision of 700 and accuracy and precision goals of 350.
The inverse of $\tilde{\mathbb{D}}^{\rm T}\tilde{\mathbb{D}}$ is computed using the built-in \textit{Mathematica} function \texttt{Inverse}, with a working precision of 300.

Recall that $H_1(r,\chi)$ is required to satisfy a vanishing boundary condition at spatial infinity.
In practice, however, due to numerical truncation errors, the spectral solution for $H_1(r,\chi)$ does not generally vanish in the limit $r\to\infty$.
As in the scalar-field case, we identify and remove the bare terms of $H_1(r,\chi)$, defined as
\begin{equation}
H_{1,\mathrm B}(N) = \lim_{r\to\infty} H_1(N),
\end{equation}
which depend only on $\chi$ and contain no inverse powers of $r$.
To enforce the correct asymptotic behavior, we subtract these bare terms from the solution at each spectral order,
\begin{equation}
H_1(N)-H_{1,\mathrm B}(N) \rightarrow H_1(N),
\end{equation}
thereby restoring the required boundary condition at spatial infinity.

\subsection{Numerical results}

\subsubsection{Backward modulus difference}

\begin{figure*}[htp!]
\centering  
\subfloat{\includegraphics[width=0.47\linewidth]{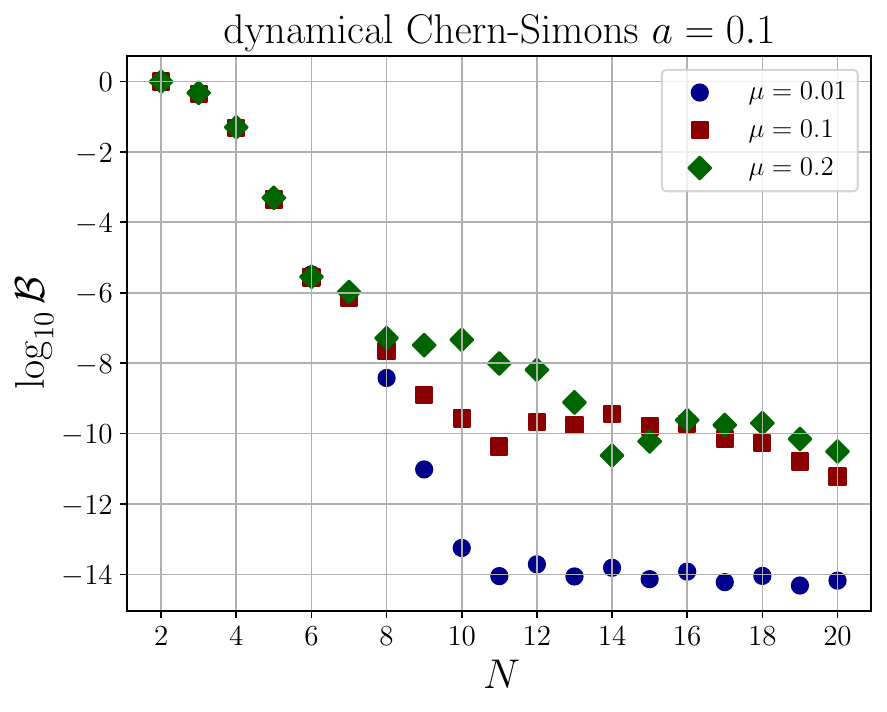}}
\subfloat{\includegraphics[width=0.47\linewidth]{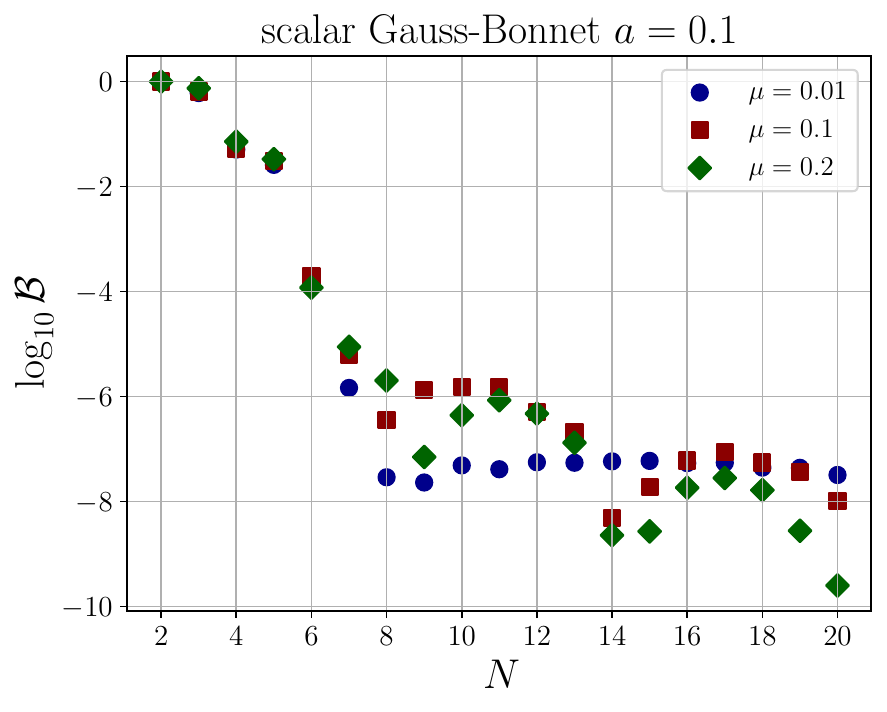}}
\caption{The backward-modulus difference [see Eq.~\eqref{eq:BMD_H} in the main text for definition] of the metric modifications due to the presence of a scalar field of mass $\mu = 0.01$ (dark-blue circles), 0.1 (dark-red squares) and 0.2 (dark-green diamonds) around a rotating black hole of dimensionless spin $a=0.1$ in dCS (left panel) and sGB (right panel) gravity as a function of the spectral order $N$. 
}
\label{fig:BMD_H_a_010}
\end{figure*}

To gauge the convergence of the metric modifications, we define the backward modulus difference $\mathcal{B}(N)$ that the changes of $H_i(r, \chi)$ computed at a given spectral $N$ from that computed at the previous spectral order $N-1$ as follows
\begin{equation}\label{eq:BMD_H}
\mathcal{B}(N) = \left[\int_{r_+}^{+\infty} \int_{-1}^{+1} \sum_{i=1}^{4} \left[ H_i (N) - H_i (N-1) \right]^2 \frac{dr}{r^2} d \chi \right]^{\frac{1}{2}}. 
\end{equation}
We include a factor of $r^{-2}$ as it is the simplest regularization factor such that $\mathcal{B}(N)$ is finite and well defined, as $ H_i $ can approach to a constant as $r \rightarrow + \infty$. 

Figure~\ref{fig:BMD_H_a_010} shows $\mathcal{B}(N)$ as a function of the spectral order $N$ for the $H_i(r, \chi)$ of rotating black holes of dimensionless spin $a=0.1$ surrounded by a massive scalar field of $\mu = 0.01$, 0.1 and 0.2 in dCS (right panel) and sGB gravity (left). 
Note that we only show the results of spectral order up to 20 because we observe nonphysical oscillations across the $\chi$-ordinate near the north and south poles for $N \geq 20$, a sign of over-resolution. 
From Fig.~\ref{fig:BMD_H_a_010}, we observed that $\mathcal{B}(N)$ decays approximately exponentially, a tendency that is similarly exhibited by $\mathcal{B}_{\vartheta} (N)$. 
Nonetheless, $\mathcal{B} (N)$ shows significantly more fluctuation from the exponential decay tendency than $\mathcal{B}_{\vartheta} (N)$. 
This is reasonable because the linear algebraic equations concerning $H_i(r, \chi)$ involves significantly more unknowns than that concerning $\varphi$. 

\subsubsection{Absolute error}

\begin{figure*}[htp!]
\centering  
\subfloat{\includegraphics[width=0.47\linewidth]{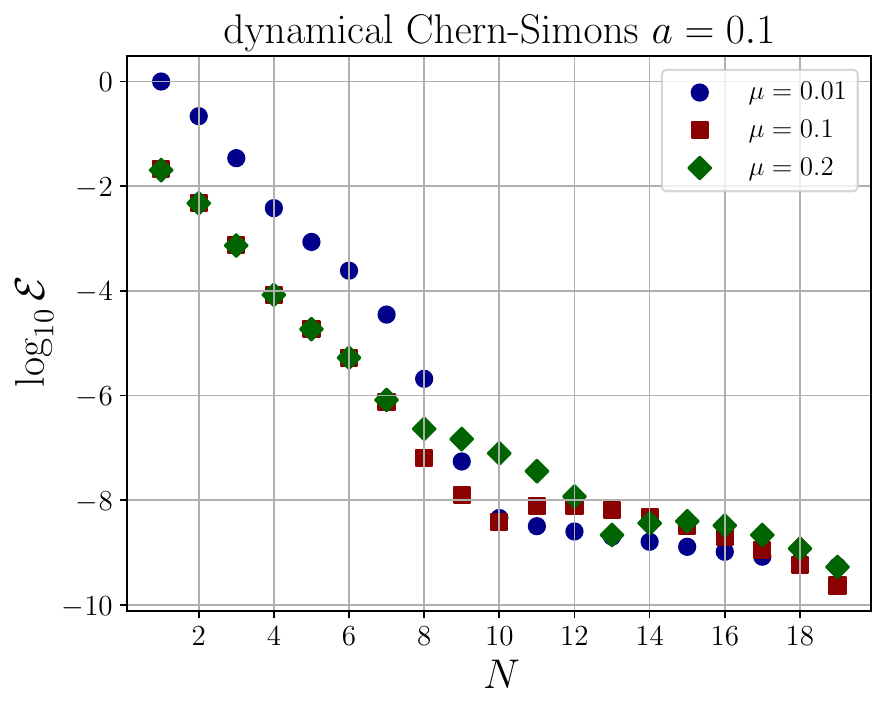}}
\subfloat{\includegraphics[width=0.47\linewidth]{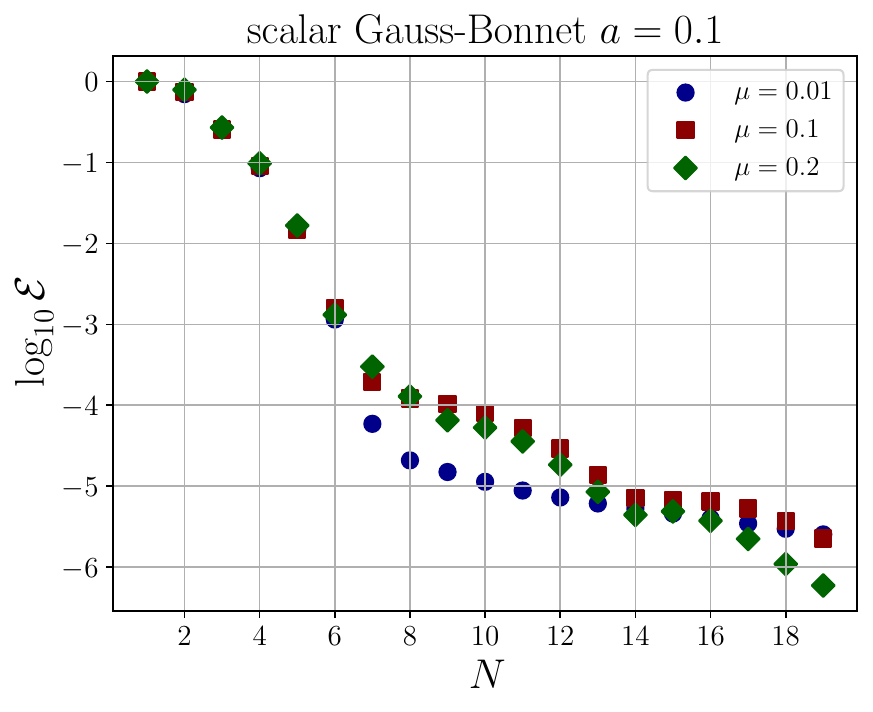}}
\caption{The error $\mathcal{E}$ [defined in Eq.~\eqref{eq:EEResidual}] of the spacetime modifications [see Eq.~\eqref{eq:HAnsatz}] for a rotating black hole with dimensionless spin $a=0.1$, sourced by a massive scalar field with $\mu=0.01$ (dark-blue circles), $\mu=0.1$ (dark-red squares), and $\mu=0.2$ (dark-green diamonds), in dCS gravity (left panel) and sGB gravity (right panel), as a function of the spectral order $N$.
Because there is no unique or unambiguous definition of a residual norm, the error is normalized such that $\mathcal{E}(N=1)=1$.
}
\label{fig:Err_EE_a_010}
\end{figure*}

To quantify how well the metric modifications $H_i$ and the scalar field $\bar{\vartheta}$ satisfy the modified Einstein equations, we define the following error: 
\begin{equation}\label{eq:EEResidual}
\mathcal{E} \propto \left[ \int_{r_+}^{\infty} \int_{-1}^{+1} E_{\beta}{}^{\nu}E_{\nu}{}^{\beta} \frac{\Delta^4}{r^{32}} (1 - \chi^2)^2 \left( -g^{(0)} \right)^{6} \,drd\chi \right]^{1/2}
\end{equation}
The factor $\Delta^{4}(1-\chi^{2})^{2}\left(-g^{(0)}\right)^{6}/r^{32}$ is the simplest regularization weight we have identified that renders the integral finite and well defined.
As in the definition of $\mathcal{E}_{\vartheta}$, the introduction of such a regularization factor is legitimate, since there is no unique or unambiguous way to define a residual norm for a system of differential equations.
Multiplying the equations by any nonvanishing function yields an equivalent system.
Accordingly, only the relative variation of $\mathcal{E}$ across spectral orders carries numerical significance.
We therefore fix the overall proportionality constant by normalizing the residual such that $\mathcal{E}(N=1)=1$.
Also as in the scalar-field case, the inclusion of this regularization factor further allows us to express the integrand in Eq.~\eqref{eq:EEResidual} in the following form,  
\begin{equation}
\begin{split}
& E_{\beta}{}^{\nu}E_{\nu}{}^{\beta} \frac{\Delta^4}{r^{32}} (1 - \chi^2)^2 \left( -g^{(0)} \right)^{6} \\
& = \sum_{j=2} \sum_{k=0} \sum_{q=0}^{4} \eta_{j,k,q} \frac{\exp(-q \mu r)}{r^{j}}\chi^k. 
\end{split}
\end{equation}
Here $\eta_{j,k,q}$ are constant coefficients that can be computed algebraically from the coefficient vector $\mathbf{v}$.
Using Eq.~\eqref{eq:math_trick}, the error $\mathcal{E}$ can therefore be evaluated analytically, up to the coefficients $\eta_{j,k,q}$ numerically determined from $v^{i}_{n \ell}$.

Figure~\ref{fig:Err_EE_a_010} shows the error $\mathcal{E}$ associated with the metric modifications around a rotating black hole with $a=0.1$, sourced by a massive scalar field with $\mu=0.01$ (blue circles), $0.1$ (red squares), and $0.2$ (green diamonds), in dCS (left panel) and sGB gravity (right panel), as a function of the spectral order $N$.
We observe that, similarly to the scalar-field residual $\mathcal{E}_{\vartheta}$, $\mathcal{E}$ initially decreases approximately exponentially with increasing $N$ up to a characteristic spectral order, beyond which it continues to decrease but at a different rate.
However, the spectral order at which this change in convergence rate occurs for $\mathcal{E}$ is approximately half of that observed for $\mathcal{E}_{\vartheta}$.
This behavior can be understood from the quadratic dependence of the stress–energy tensor $\tilde{T}_{\beta}{}^{\nu}$, part of the source terms in the modified Einstein equations, on the scalar field.
Since $\vartheta \sim e^{-\mu r}$, it follows that $\tilde{T}_{\beta}{}^{\nu} \sim e^{-2\mu r}$, which introduces radial variations that are effectively twice as ``stiff" as those of the scalar field itself.
As a consequence, the metric modifications reach the over-resolution regime at a spectral order that is approximately a factor of 2 smaller than that required for the scalar field.

As in the case of the scalar fields, to minimize the error that would propagate to subsequent computations, we select $H_i(r, \chi)$ that minimize $\mathcal{E}$ by reading the optimal spectral order as follows: 
\begin{equation}\label{eq:opt_spec_order_H} 
N_{\rm opt} = \text{arg}\min_{N} \mathcal{E}. 
\end{equation}

\begin{figure*}[htp!]
\centering  
\subfloat{\includegraphics[width=0.47\linewidth]{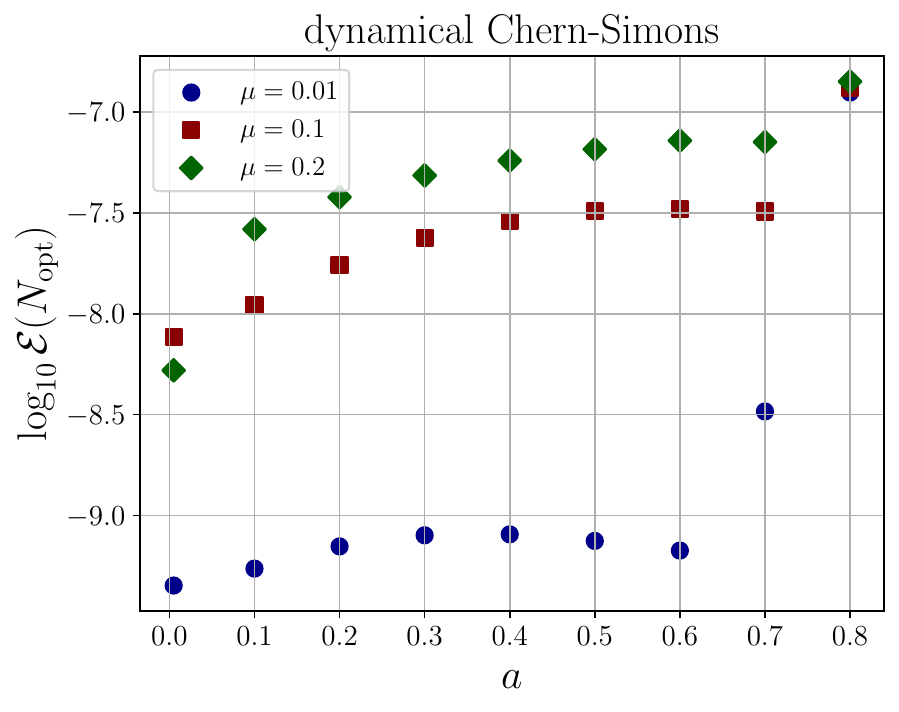}}
\subfloat{\includegraphics[width=0.47\linewidth]{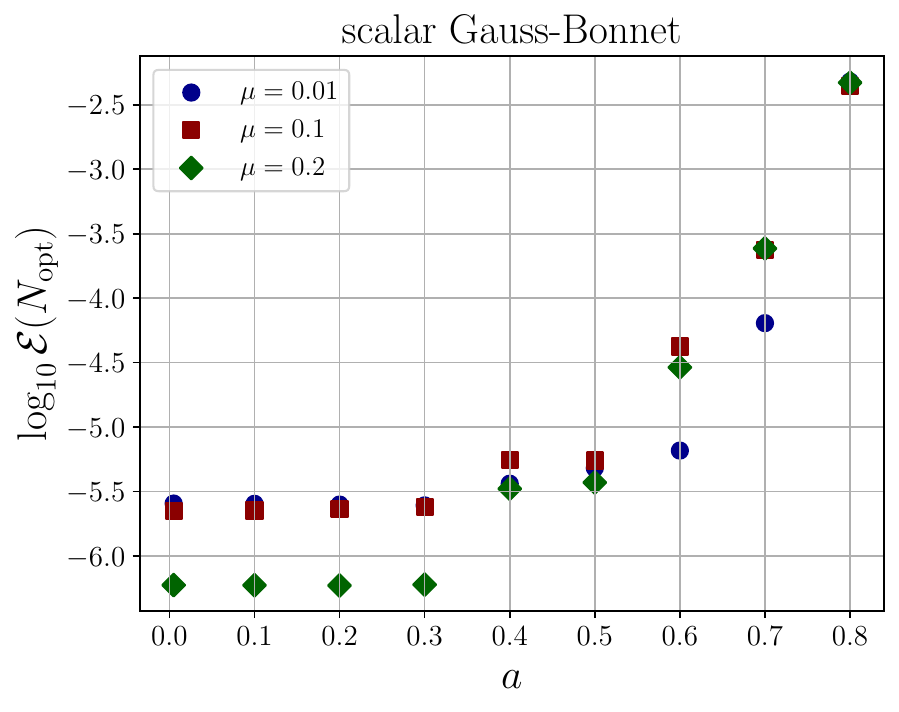}}
\caption{The least residual of the modified Einstein equations [see Eq.~\eqref{eq:Err_scalar} for the definition] for a rotating black hole surrounded by a massive scalar field of mass $\mu = 0.01$ (dark-blue circles), $\mu = 0.1$ (dark-red squares), and $\mu = 0.2$ (dark-green diamonds), shown as a function of the dimensionless spin $a$.
The left panel corresponds to dCS gravity, and the right to sGB gravity.
Since there is no unique, unambiguous definition of an equation residual, the error is normalized such that its value at spectral order $N=1$ is unity for all $a$ and $\mu$.
}
\label{fig:Min_err_EE_a}
\end{figure*}

Figure~\ref{fig:Min_err_EE_a} shows the least error, $\mathcal{E}(N_{\rm opt})$, as a function of the dimensionless spin $a$ for $\mu=0.01$, $0.1$, and $0.2$ in dCS (left panel) and sGB (right panel) gravity.
$H_i(r, \chi)$ in axi-dilaton gravity could be obtained by summing the modifications in dCS and sGB gravity. 
In contrast to the behavior of $\mathcal{E}_{\vartheta}(N{\rm opt})$, we find that $\mathcal{E}(N_{\rm opt})$ increases systematically with spin, reaching values of order $\sim 10^{-3}$ at $a=0.8$, while remaining at a constant approximately as $\mu$ varies from $0.01$ to $0.2$.
This trend is expected, as the structure of rotating black-hole spacetimes becomes increasingly complicated at higher spin, a behavior also observed in Ref.~\cite{Lam:2025fzi}.
For this reason, we restrict our analysis to spins $a\leq 0.8$, beyond which the present spectral implementation does not achieve sufficient accuracy for constructing black-hole spacetimes with massive scalar charges.

\begin{figure*}[htp!]
\centering  
\subfloat{\includegraphics[width=0.31\linewidth]{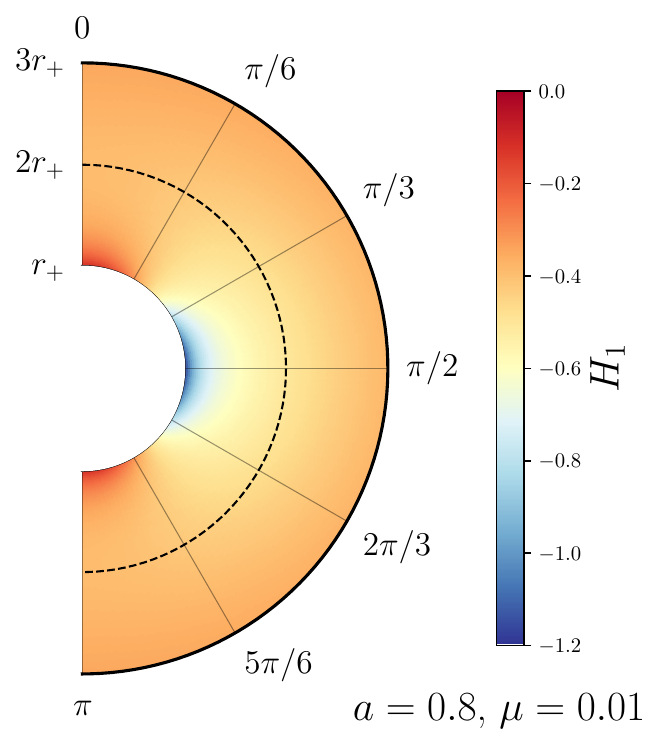}}
\subfloat{\includegraphics[width=0.31\linewidth]{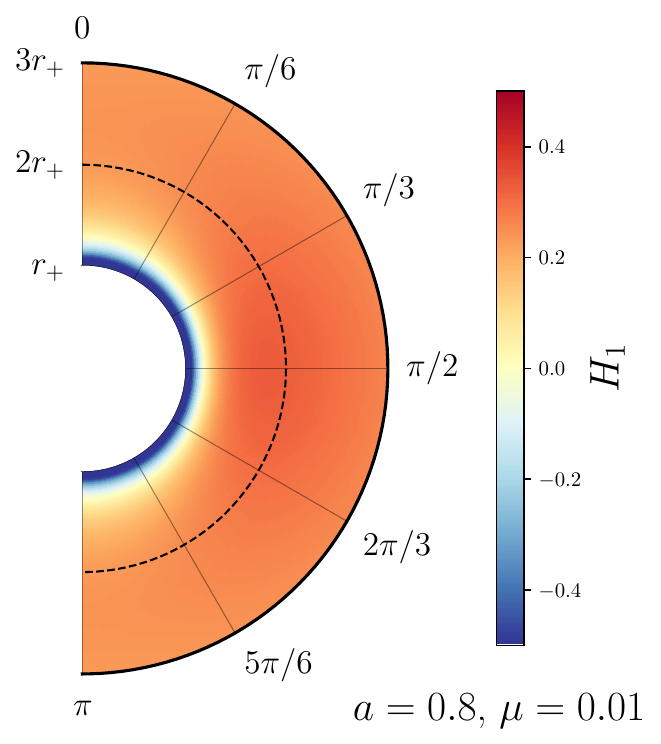}}
\qquad
\subfloat{\includegraphics[width=0.31\linewidth]{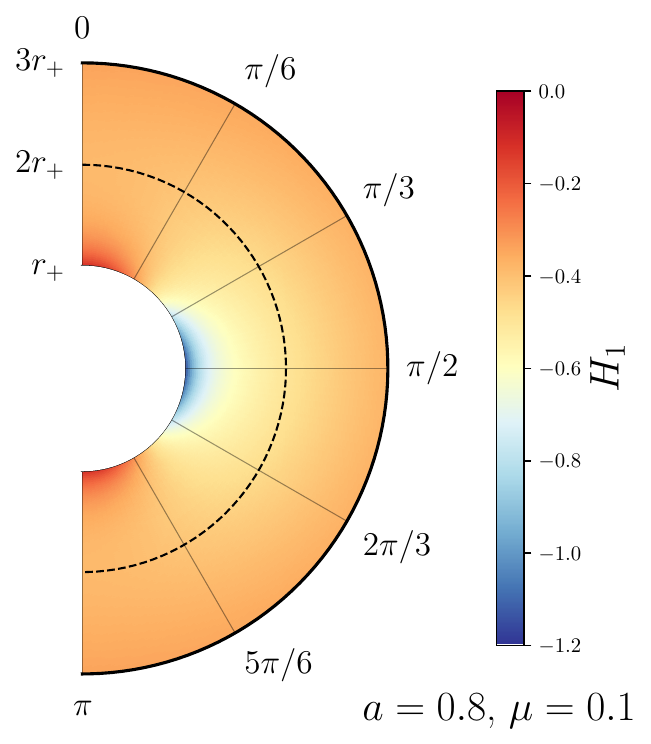}}
\subfloat{\includegraphics[width=0.31\linewidth]{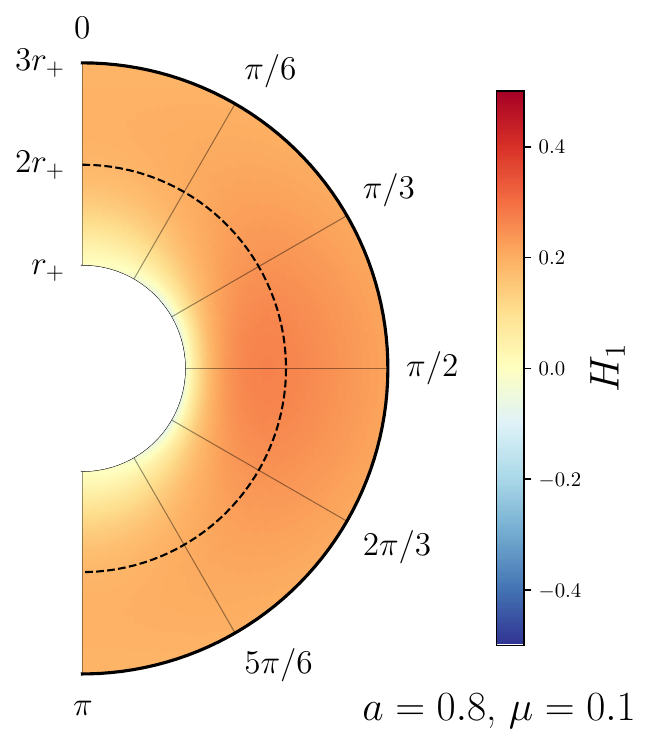}}
\qquad
\subfloat{\includegraphics[width=0.31\linewidth]{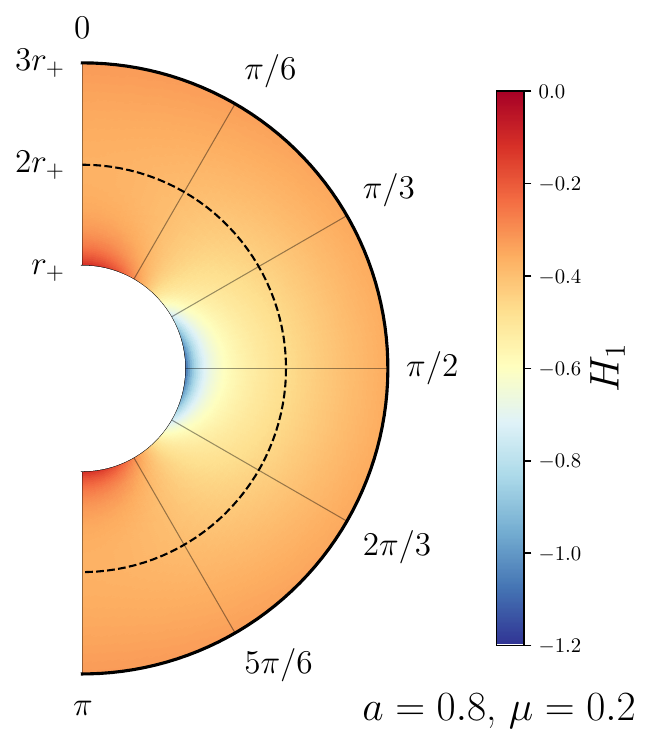}}
\subfloat{\includegraphics[width=0.31\linewidth]{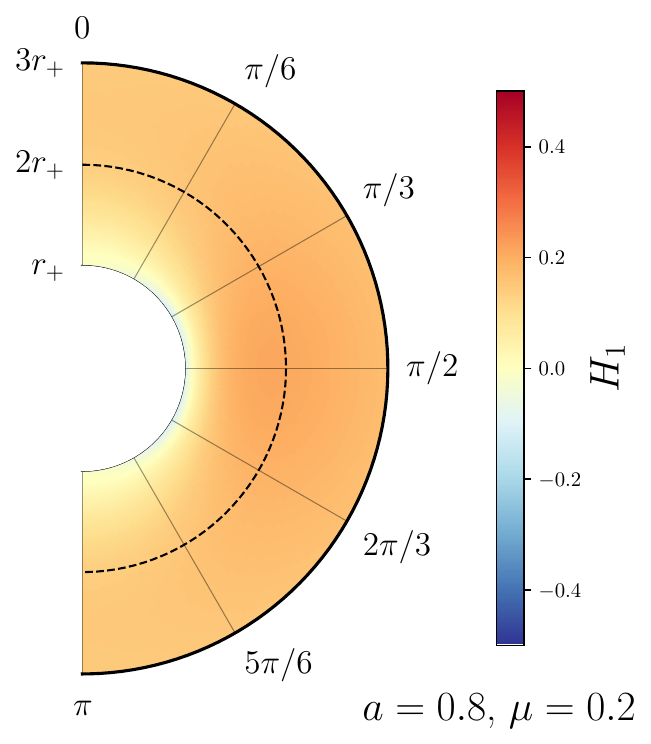}}
\caption{Meridional cross-sections of $H_1$ of a rotating black hole with dimensionless spin $a = 0.8$ in dCS (left panels) and sGB (right panels) gravity. 
The top, middle, and bottom panels correspond to scalar-field masses $\mu = 0.01$, $\mu = 0.1$, and $\mu = 0.2$, respectively. 
All solutions are computed at their respective optimal spectral resolutions.  
By comparing with the results in \cite{Lam:2025fzi}, the overall multipolar structure of $H_1$ is not significantly altered by the mass of the scalar fields.
}
\label{fig:H1_meridian}
\end{figure*}

\begin{figure*}[htp!]
\centering  
\subfloat{\includegraphics[width=0.31\linewidth]{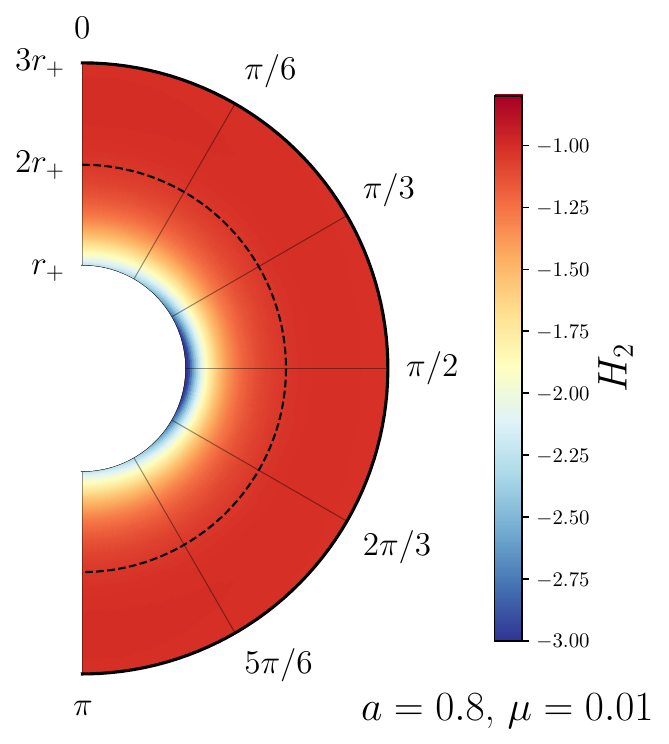}}
\subfloat{\includegraphics[width=0.31\linewidth]{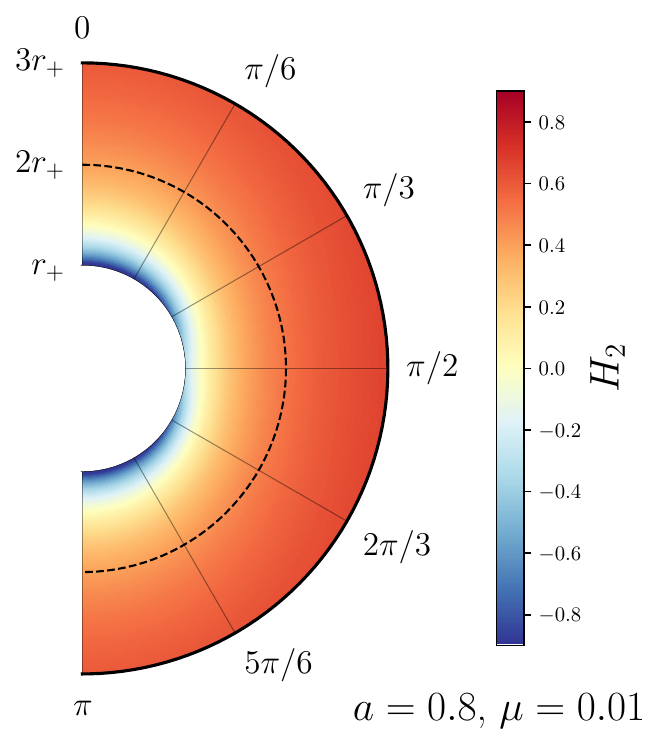}}
\qquad
\subfloat{\includegraphics[width=0.31\linewidth]{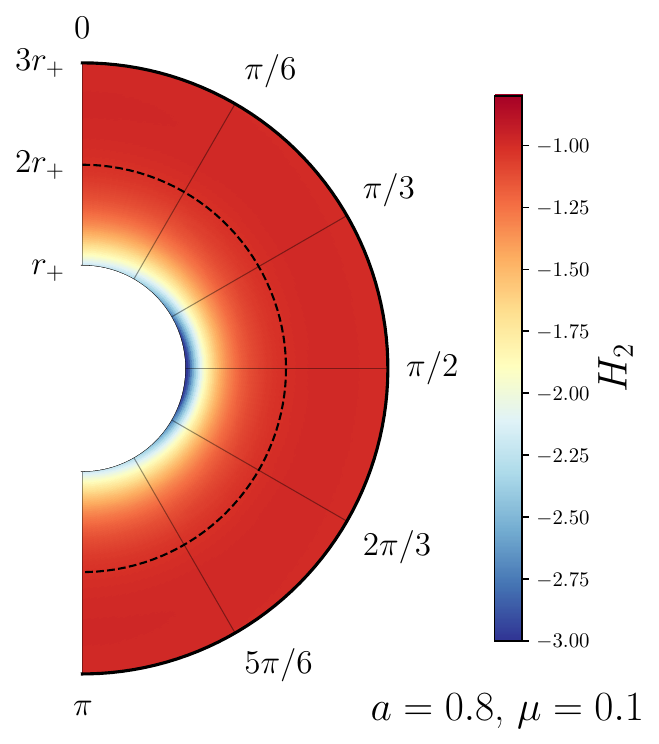}}
\subfloat{\includegraphics[width=0.31\linewidth]{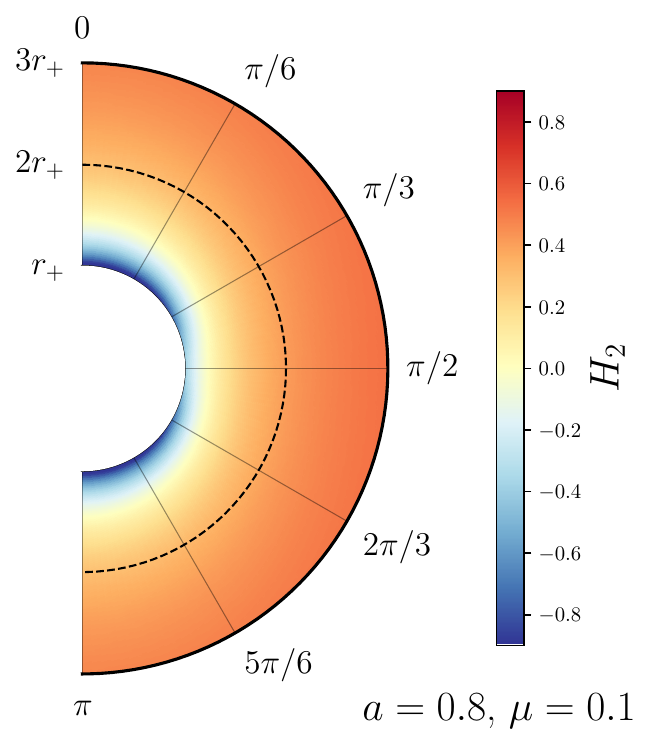}}
\qquad
\subfloat{\includegraphics[width=0.31\linewidth]{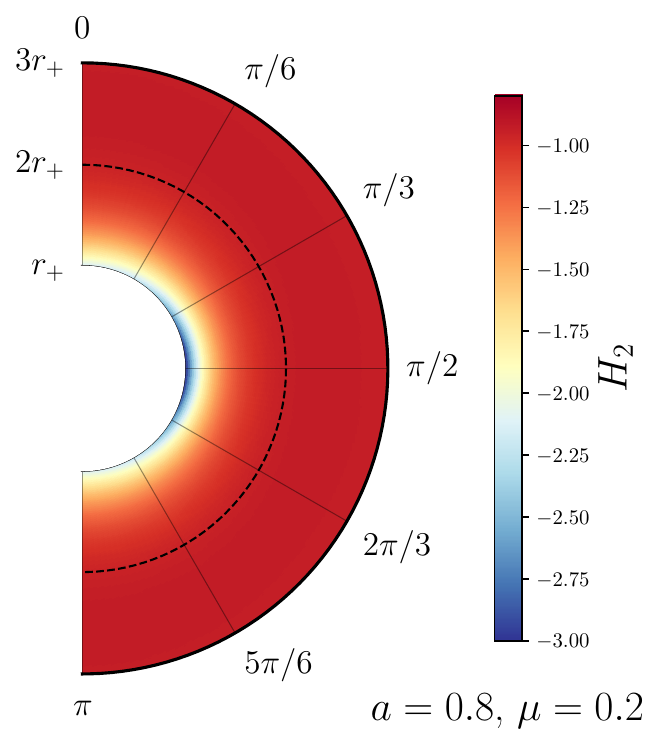}}
\subfloat{\includegraphics[width=0.31\linewidth]{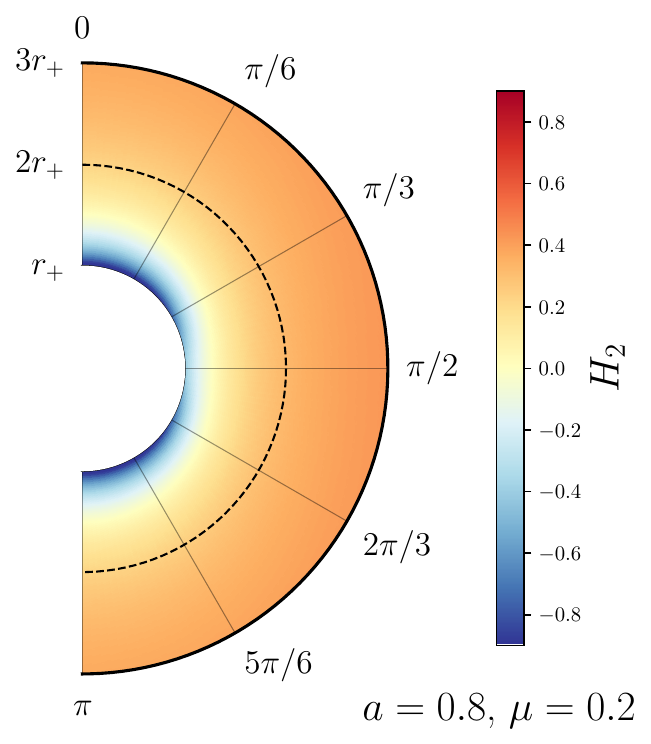}}
\caption{Identical to Fig.~\ref{fig:H1_meridian}, except that $H_2$ is shown. 
}
\label{fig:H2_meridian}
\end{figure*}

Figsures~\ref{fig:H1_meridian} and~\ref{fig:H2_meridian} show the meridional cross-sections of the metric modifications for a rotating black hole with dimensionless spin $a=0.8$, surrounded by a massive scalar field with $\mu=0.01$ (top), $0.1$ (middle), and $0.2$ (bottom), in dCS (left panels) and sGB (right panels) gravity.
For clarity and brevity, we show only $H_1$ (Fig.~\ref{fig:H1_meridian}) and $H_2$ (Fig.~\ref{fig:H2_meridian}), which correspond to corrections to the $tt$ and $t\phi$ components of the metric, respectively.
The metric corrections of are computed using the corresponding optimal spectral order [selected according to Eq.~\eqref{eq:opt_spec_order_H}] at the given $a$ and $\mu$. 
These components are related to the lapse function and shift vector experienced by stationary observers at fixed spatial locations.
From Figs.~\ref{fig:H1_meridian} and~\ref{fig:H2_meridian}, we first observe that $H_1$ and $H_2$ are even in $\chi$, as expected. 
Next, we observed that, as in the case of scalar field (cf.~Fig.~\ref{fig:Scalar_meridian}), the scalar-field mass does not significantly alter the geometric structure of the metric corrections.
Instead, increasing the scalar-field mass primarily reduces the overall magnitude of the spacetime deformations.
This behavior can be explained by the more rapid spatial decay of heavier scalar fields, which results in a smaller effective scalar-field energy density available to source the metric modifications.

%%%%%%%%%%%%%%%%%%%%%%%%%%%%%%%%%%%%%%%%%%%%%%%%%%%%%%%%%%%%%%%%%%%%%%%%%%%%%%%%%%%%%%%%%%%%%%%%%%%%%%%%%%%%%%%%%%%%%%%%%%%%
\section{Physical properties of the spacetime} \label{sec:Physical_properties}

In this section, we compute the leading-order-in-$\zeta$ corrections to the horizon angular velocity and the surface gravity of rotating black holes surrounded by a massive scalar field. 
These corrections are defined as the difference between the modified quantities and their general-relativistic values. 
% Both quantities are valid up to only the leading order in $\zeta$.  
As in the case of $H_i(r, \chi)$, the leading-order modifications to these quantities in axi-dilaton gravity is just the sum of the modification in dCS and sGB gravity. 
Thus, we shall just focusing on computing the modifications in these two gravity theories. 
Both quantities are essential ingredients for the computation of black-hole quasinormal-mode frequencies \cite{Chung:2023wkd, Chung:2024vaf, Chung:2025gyg}, which might lead to search for massive scalar charges via gravitational-wave detections. 

\subsection{Horizon angular velocity}

We first compute the angular velocity, $ \Omega_{H}$, of the event horizon of rotating black holes surrounded by a massive scalar charges in dCS and sGB gravity 
Formally, the horizon angular velocity is given by 
\begin{equation}\label{eq:Omega_H_def}
\Omega_{H} = \frac{g_{t \phi}}{g_{\phi \phi}} \Bigg|_{r=r_+}. 
\end{equation}
The leading-order modification to the horizon angular velocity, $\Omega_{H}^{(1)}$, could then be computed from $H_i(r, \chi)$ via the following expression \cite{Cano_Ruiperez_2019, Lam:2025fzi}, 
\begin{equation}\label{eq:Omega_H_1}
\Omega_H^{(1)} = \frac{a}{2Mr_+}\Big(H_2 - H_4\Big)\Big|_{r=r_+}.
\end{equation}

\begin{figure*}[htp!]
\centering  
\subfloat{\includegraphics[width=0.47\linewidth]{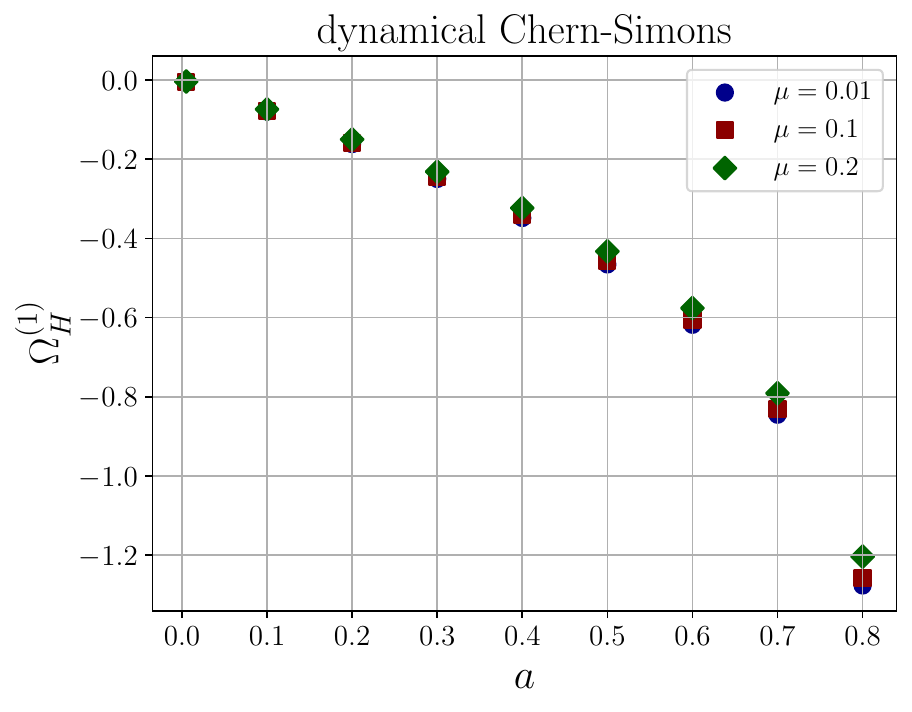}}
\subfloat{\includegraphics[width=0.47\linewidth]{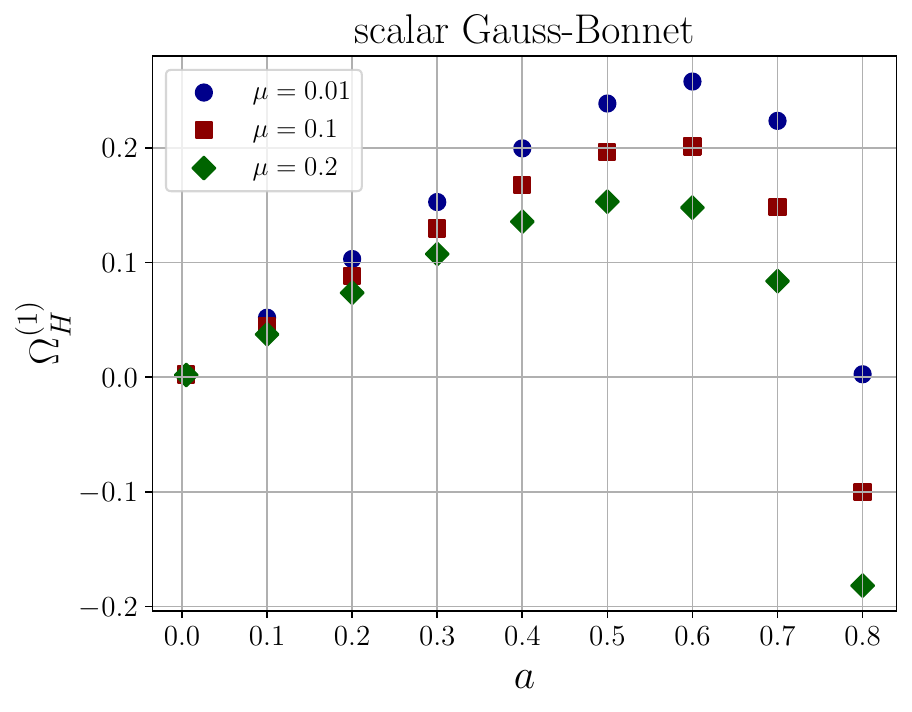}}
\caption{The leading-order modification to the angular velocity of the event horizon, $\Omega_{H}^{(1)}$ [see Eqs.~\eqref{eq:Omega_H_def} and \eqref{eq:Omega_H_1} for definition], of a rotating black hole surrounded by a scalar field of mass $\mu = 0.01$ (circles in dark blue), 0.1 (squares in dark red) and 0.2 (diamonds in dark green) in dCS (left panel) and sGB (right panel) gravity as a function of $a$.  
}
\label{fig:Omega_H}
\end{figure*}

Figure~\ref{fig:Omega_H} shows $\Omega_{H}^{(1)}$ for rotating black holes surrounded by a scalar field of $\mu = 0.01$ (blue circles), $0.1$ (red squares) and $0.2$ (green diamonds) subjected to dCS (left panel) and sGB (right panel) as a function of $a$. 
To minimize the error, $\Omega_{H}^{(1)}$ is computed using $H_i(r, \chi)$ at the optimal spectral order at the corresponding $a$ and $\mu$, as selected according to Eq.~\eqref{eq:opt_spec_order_H}. 
We observe that the presence of the massive scalar field tends to decrease the horizon angular velocity of black holes in dCS gravity for dimensionless spin of $a \leq 0.8$; while first increase $\Omega_{\rm H}^{(1)}$ for black holes in sGB gravity of dimensionless spin of $a \lesssim 0.7$, and then decrease $\Omega_{\rm H}^{(1)}$ from that dimensionless spin onward. 
Both tendencies are consistent with that in the case of the massless scalar field in the corresponding gravity theory (see \cite{Lam:2025fzi}).

\begin{figure*}[htp!]
\centering  
\subfloat{\includegraphics[width=0.47\linewidth]{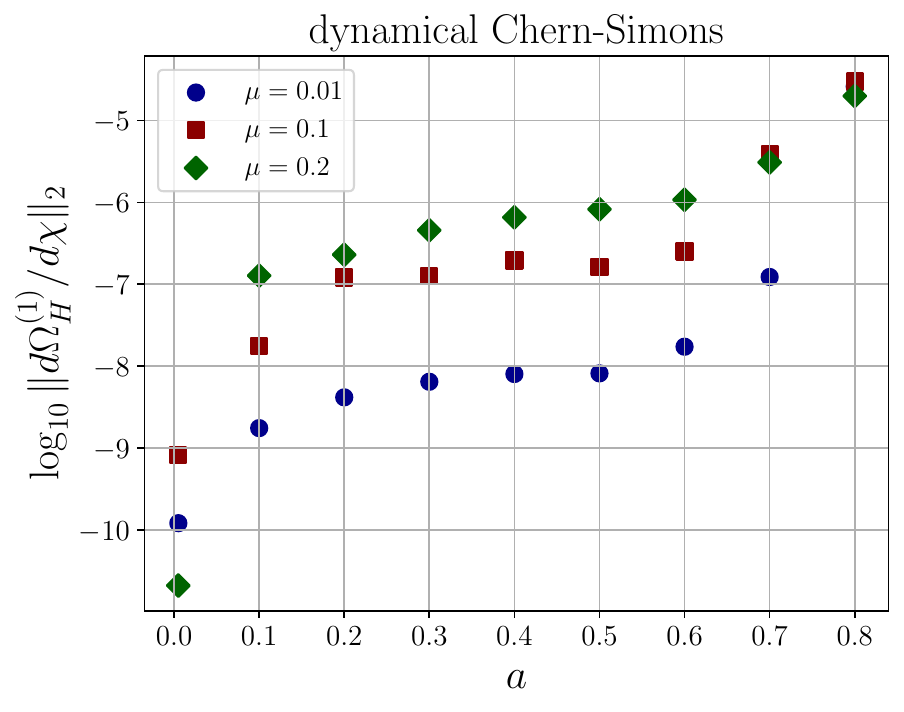}}
\subfloat{\includegraphics[width=0.47\linewidth]{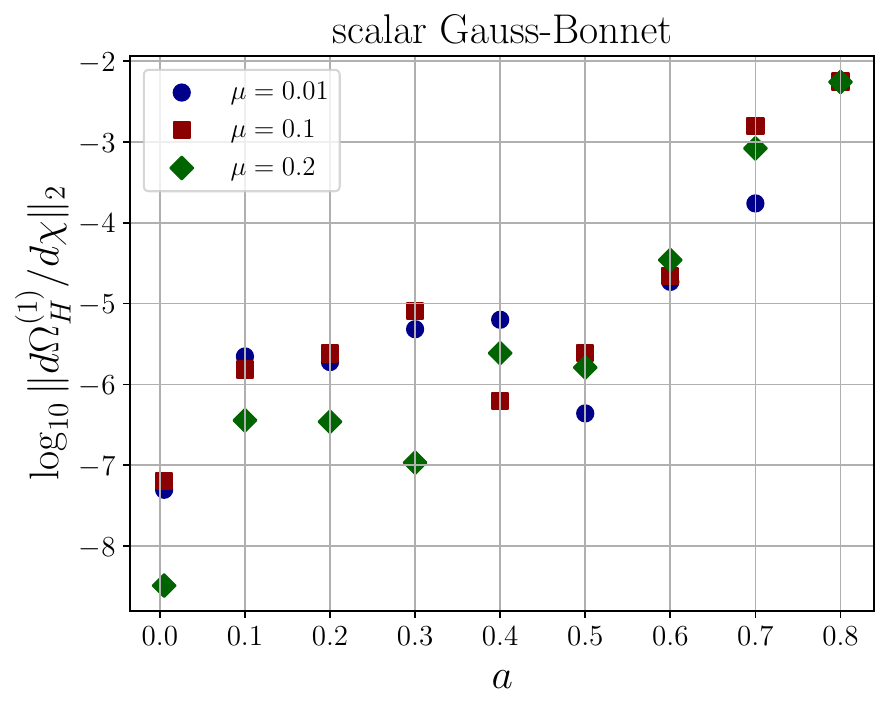}}
\caption{$\left\|d\Omega_H^{(1)}/d\chi\right\|_2$ [see Eq.~\eqref{eq:HorizonAngularVelocityChiError} for definition] of a rotating black hole surrounded by a scalar field of mass $\mu = 0.01$ (circles in dark blue), 0.1 (squares in dark red) and 0.2 (diamonds in dark green) in dCS (left panel) and sGB (right panel) gravity as a function of $a$.  
}
\label{fig:d_Omega_H}
\end{figure*}

Following Ref.~\cite{Lam:2025fzi}, we compute the $L^{2}$ norm of the $\chi$ derivative of $\Omega_{H}^{(1)}$ as a consistency check of both our numerical evaluation of $\Omega_{H}^{(1)}$ and the validity of the adopted metric ansatz,
\begin{equation}\label{eq:HorizonAngularVelocityChiError}
\left|\!\left|\frac{d\Omega_H^{(1)}}{d\chi}\right|\!\right|_2 = \left[ \int_{-1}^{+1} \left(\frac{d\Omega_H^{(1)}}{d{\chi}} \right)^2 \,d\chi \right]^{1/2}. 
\end{equation}
According to black-hole mechanics \cite{Poisson:2009pwt, Bardeen:1973gs}, the angular velocity of the Killing horizon of a stationary and axisymmetric black hole must be constant over the entire horizon, which is a sphere in Boyer–Lindquist coordinates.
Therefore, an exact solution would satisfy $\left|\!\left|d\Omega_H^{(1)}/d\chi\right|\!\right|_2 = 0$.
Hence, the magnitude of this $L^{2}$ norm provides a measure of the error in our leading-order spacetime modification, as well as a diagnostic of any inconsistency between the computed solution and the assumption that the horizon remains a Killing horizon.
Figure~\ref{fig:d_Omega_H} shows $\left|\!\left|d\Omega_H^{(1)}/d\chi\right|\!\right|_2 $ for black holes surrounded by massive scalar fields with $\mu = 0.01$ (blue circles), $0.1$ (red squares), and $0.2$ (green diamons) in dCS (left panel) and sGB (right panel) gravity.
For both gravity theories and all scalar masses considered, we find that $\left|\!\left|d\Omega_H^{(1)}/d\chi\right|\!\right|_2 $ increases monotonically with the dimensionless spin $a$.
This behavior is consistent with the results reported in Ref.~\cite{Lam:2025fzi} and reflects the increasing complexity of rotating black-hole spacetimes at higher spin.
Importantly, even in the least accurate cases, we find $\left|\!\left|d\Omega_H^{(1)}/d\chi\right|\!\right|_2 \lesssim 10^{-2}$, corresponding to at most a $\sim 10\%$ variation relative to $\Omega_H^{(1)}$.
This demonstrates that the spacetime modifications constructed using our spectral methods remain accurate up to $a = 0.8$.

\subsection{Surface gravity}

Next, we compute the changes to the surface gravity, $\kappa$, of black holes surrounded by massive scalar charges. 
Given a black hole that has a Killing horizon, $\kappa$ can be computed through constructing a timelike Killing vector, and then consider the inner product of the derivatives of the Killing vector. 
The analytical expression for computing $\kappa$ through this way is available as, for example, Eq.~(67) of \cite{Lam:2025fzi} or \cite{Poisson:2009pwt}, and we shall not reproduce here. 
Naturally, $\kappa$ is nonlinear in $\zeta$ and $H_i(r, \chi)$, but the leading-order modification to the surface gravity, $\kappa^{(1)}$, can be computed via the following expression \cite{Cano_Ruiperez_2019, Lam:2025fzi}: 
\begin{equation}\label{eq:SurfaceGravity}
\begin{split}
\kappa^{(1)} = \frac{r_+ - M}{2Mr_+} \bigg[ &  H_2 - \frac{H_3}{2} - \frac{H_4}{2} + \frac{M^2 r_+^2}{(r_+ - M)\Sigma} \\
& \frac{d}{dr}\left(-H_1 \Sigma + a^2(1-\chi^2)(2H_2 - H_4)\right) \\
&+ 2(r_+ - M)(H_4 - 2H_2) \bigg] \bigg|_{r=r_+}. 
\end{split}
\end{equation}

\begin{figure*}[htp!]
\centering  
\subfloat{\includegraphics[width=0.47\linewidth]{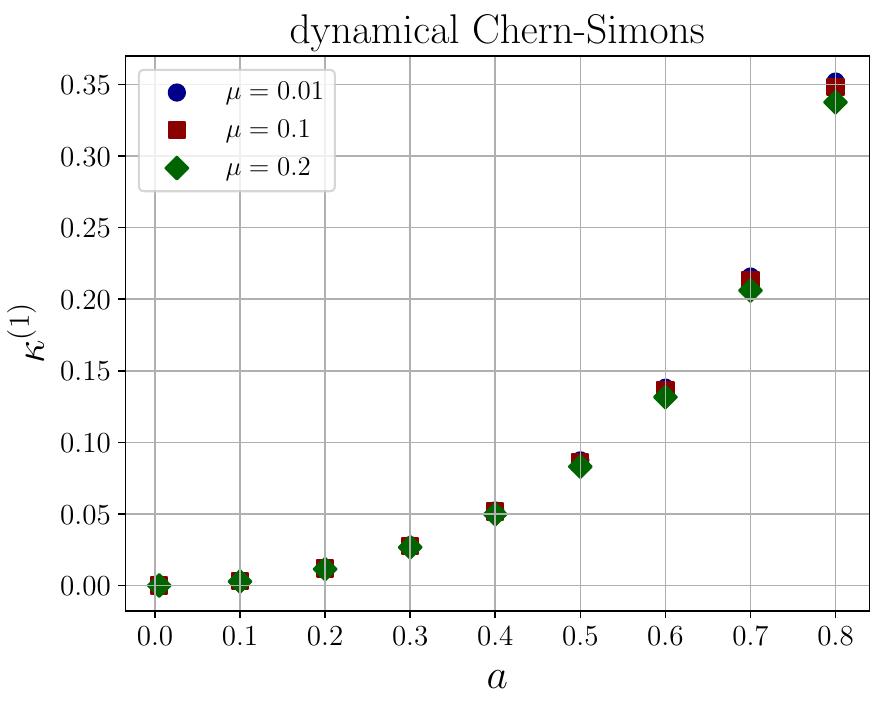}}
\subfloat{\includegraphics[width=0.47\linewidth]{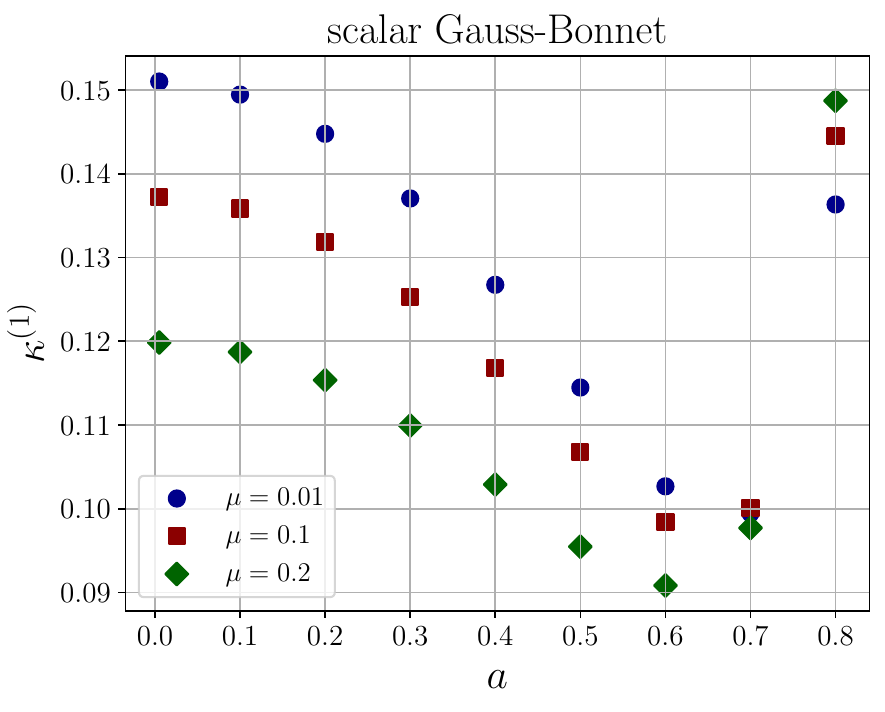}}
\caption{Identification to Fig.~\ref{fig:Omega_H}, except that the leading-order modification to the surface gravity is shown here. 
}
\label{fig:kappa}
\end{figure*}

Figure~\ref{fig:kappa} shows the leading-order correction to the surface gravity, $\kappa^{(1)}$, for rotating black holes surrounded by massive scalar fields with $\mu = 0.01$ (blue circles), $0.1$ (red squares), and $0.2$ (green diamonds), in dCS (left panel) and sGB (right panel) gravity, as a function of the dimensionless spin $a$.
As in the computation of the horizon angular velocity, $\kappa^{(1)}$ is evaluated using the metric functions $H_i(r,\chi)$ at the optimal spectral order corresponding to each pair $(a,\mu)$, selected according to Eq.~\eqref{eq:opt_spec_order_H} to minimize numerical error.
We find that, in both gravity theories, the presence of a massive scalar field generally increases the surface gravity relative to the general-relativistic value.
In particular, in the nonrotating limit $a\to 0$, we obtain $\kappa^{(1)} \simeq 0$ for black holes in dCS gravity, whereas $\kappa^{(1)} \sim 10^{-1}$ for black holes in sGB gravity.
This behavior is consistent with the fact that, in the nonrotating $(a\to 0)$ and massless-field $(\mu\to 0)$ limits, the Schwarzschild metric remains a solution of the modified Einstein equations in dCS. \footnote{The Schwarzschild metric is still a solution in sGB gravity. However, the Schwarzschild metric is a solution in sGB gravity without endowing a scalar hair. 
A non-rotating black hole with a non-trivial scalar hair in sGB gravity is different from the Schwarzschild metric.}
As the spin increases, $\kappa^{(1)}$ in dCS gravity grows monotonically for all scalar masses considered.
By contrast, in sGB gravity, $\kappa^{(1)}$ increases from $a=0$ up to $a\simeq 0.7$, beyond which it decreases at higher spin.
Both trends are consistent with those found in the massless-scalar limit for the corresponding theories~\cite{Lam:2025fzi}.

\begin{figure*}[htp!]
\centering  
\subfloat{\includegraphics[width=0.47\linewidth]{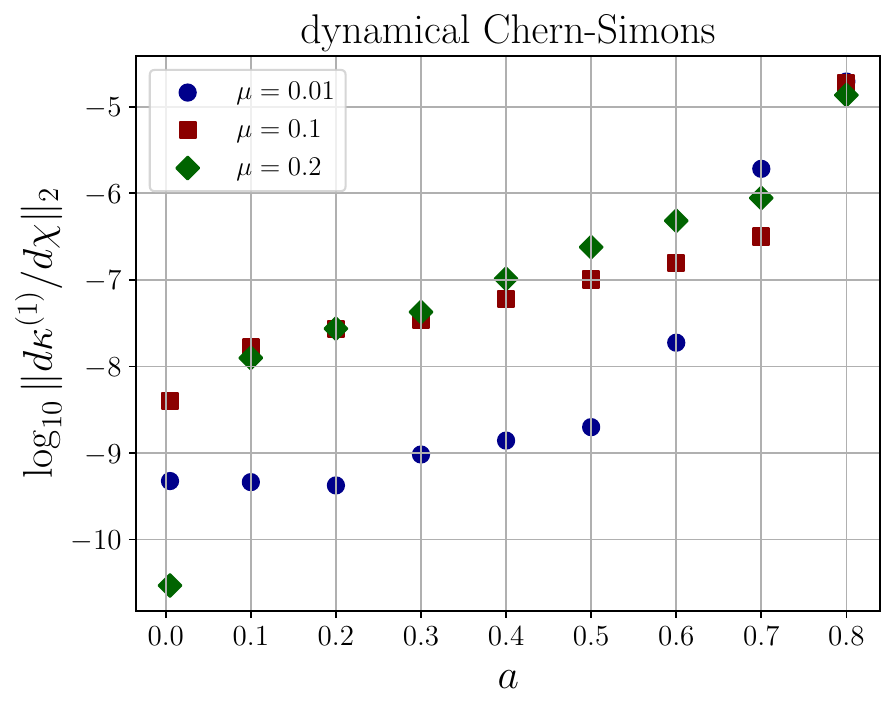}}
\subfloat{\includegraphics[width=0.47\linewidth]{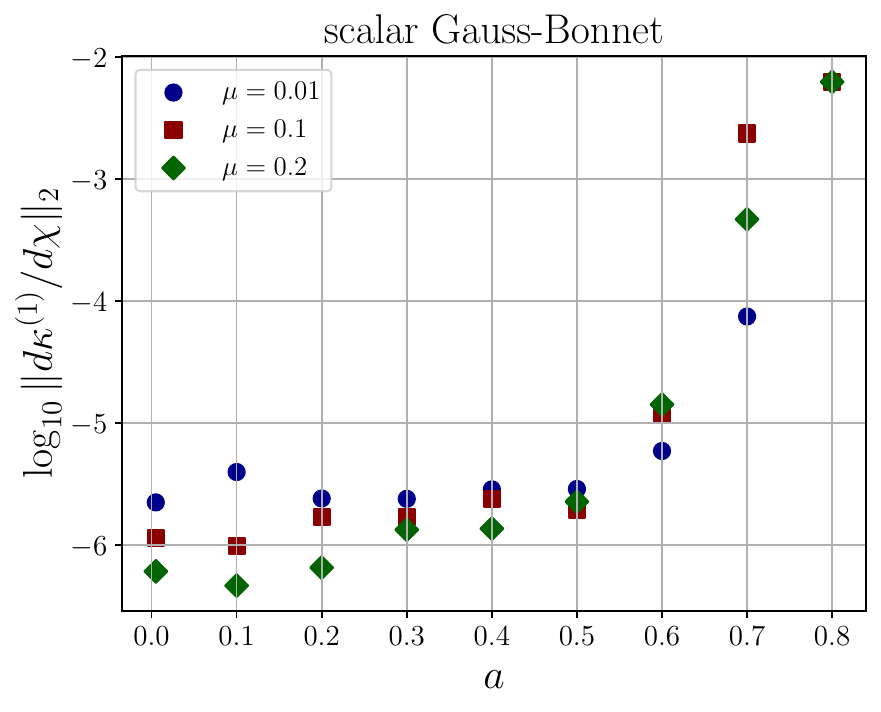}}
\caption{Identification to Fig.~\ref{fig:d_Omega_H}, except that $\left|\!\left| d\kappa^{(1)}/d\chi\right|\!\right|_2$ [see Eq.~\eqref{eq:KappaChiError} for definition] is shown here. 
}
\label{fig:d_Kappa_H}
\end{figure*}

The surface gravity is another quantity that should be constant over the event horizon according to black-hole mechanics \cite{Poisson:2009pwt, Bardeen:1973gs}.
Equivalently, the derivative $d\kappa^{(1)}/d\chi$ should vanish on the horizon.
Motivated by this requirement and following Ref.~\cite{Lam:2025fzi}, we compute the $L^{2}$ norm of the $\chi$ derivative of $\kappa^{(1)}$ as a consistency check,
\begin{equation}\label{eq:KappaChiError}
\left|\!\left|\frac{d\kappa^{(1)}}{d\chi}\right|\!\right|_2 = \left[ \int_{-1}^{+1} \left(\frac{d\kappa^{(1)}}{d{\chi}} \right)^2 \,d\chi \right]^{1/2}. 
\end{equation}
Figure~\ref{fig:d_Kappa_H} shows $\left|\!\left| d\kappa^{(1)}/d\chi\right|\!\right|_2$ for black holes surrounded by massive scalar fields with $\mu = 0.01$ (blue circles), $0.1$ (red squares), and $0.2$ (green diamonds), in dCS (left panel) and sGB (right panel) gravity.
As in the case of $\left|\!\left|d\Omega_H^{(1)}/d\chi\right|\!\right|_2$, we find that $\left|\!\left| d\kappa^{(1)}/d\chi\right|\!\right|_2$ increases monotonically with the dimensionless spin $a$ for all scalar masses and in both gravity theories.
Moreover, for fixed $(a,\mu)$, the magnitudes of $\left|\!\left| d\kappa^{(1)}/d\chi\right|\!\right|_2$ and $\left|\!\left|d\Omega_H^{(1)}/d\chi\right|\!\right|_2$ are comparable (cf.~Fig.~\ref{fig:d_Omega_H}).
We find that the maximum value of $\left|\!\left| d\kappa^{(1)}/d\chi\right|\!\right|_2 \lesssim 10^{-1}$, corresponding to at most a $\sim 10\%$ variation relative to $\kappa^{(1)}$.
This provides an additional validation of the accuracy and internal consistency of our numerically constructed spacetimes.

%%%%%%%%%%%%%%%%%%%%%%%%%%%%%%%%%%%%%%%%%%%%%%%%%%%%%%%%%%%%%%%%%%%%%%%%%%%%%%%%%%%%%%%%%%%%%%%%%%%%%%%%%%%%%%%%%%%%%%%%%%%%
\section{Concluding remarks}
\label{sec:Conclusions}

In this paper, we have developed spectral methods capable of accurately constructing the spacetime of a rotating black hole surrounded by massive scalar fields subject to nonminimal couplings to spacetime curvature.
In particular, our scheme accurately resolves massive scalar fields with masses up to $\mu \leq 0.2,M$ in dCS and sGB gravity, achieving errors $\lesssim 10^{-5}$, and yields the corresponding leading-order metric modifications with error $\lesssim 10^{-4}$ for black holes with dimensionless spins up to $a=0.8$.
Using these self-consistent spacetime solutions, we have computed a range of physical quantities that are potentially accessible through astrophysical observations.
Moreover, since the Lagrangian density of axi-dilaton gravity is given by the sum of the dynamical Chern-Simons and scalar Gauss-Bonnet contributions, the corresponding scalar fields and spacetime modifications in axi-dilaton gravity can be obtained straightforwardly by combining the results from the two theories.
Our work therefore extends the reach of existing spectral approaches \cite{Lam:2025elw,Lam:2025fzi,Fernandes:2025vxg} to gravitational systems with massive degrees of freedom.

We have also investigated the impact of the scalar-field mass on the scalar configuration, spacetime modifications, and derived physical observables, and compared our findings with those obtained in the massless case \cite{Lam:2025elw,Lam:2025fzi}.
Within the parameter space explored in this study, we find that introducing a finite scalar-field mass does not significantly alter the geometry of the scalar field or the spacetime modifications, but it does lead to changes in physical observables.
These differences may open new avenues for probing massive scalar fields, well-motivated candidates for dark matter or dark energy, through astronomical observations.
For instance, the metric modifications computed here can be incorporated into waveform models for extreme-mass-ratio inspirals, enabling future space-based gravitational-wave detectors such as LISA \cite{Maselli:2021men, Barsanti:2022vvl, Maselli:2020zgv} to search for signatures of new fundamental fields.
The results reported in this work enhance our understanding of rotating black-hole spacetimes with matter.  

The scalar fields and black-hole spacetimes constructed in this work also pave the way to direct searches for massive scalar fields from detected gravitational-wave signals.
One possible way is to use the metric-modifications to compute the effects to the inspiral dynamics of black holes with the massive scalar fields via post-Newtonian and post-Minkowskian formalisms \cite{Yagi:2012ya}. 
The impacts on the inspiral dynamics could be used to extend the search in \cite{Xie:2024xex} to massive scalar fields in dynamical Chern-Simons and axi-dilaton gravity. 
Another possibility is to search for massive scalar charges from black-hole ringdown spectroscopy.
To this end, we are currently computing the quasinormal-mode spectra of rotating black holes surrounded by massive scalar fields using METRICS (Metric pErTuRbations wIth speCtral methodS) \cite{Chung:2023wkd, Chung:2024vaf, Chung:2024ira, Chung:2025gyg}, and we will report these results in a forthcoming publication.

Overall, the methods presented here represent a significant step forward in constructing black-hole spacetimes with massive degrees of freedom.
Nevertheless, the method requires further refinement for accurately construction of scalar fields with larger masses and the associated spacetime modifications for black holes with higher spin.
One possible refinement is to analytically project the source term $\Sigma \mathscr{Q}$ onto a basis of associated Legendre polynomials, following the approach of Refs.~\cite{McNees:2015srl,Stein:2014wza}.
Such a projection would reduce the partial differential equations to a coupled system of ordinary differential equations for the radial modes $\varphi_\ell(r)$, yielding a coefficient matrix with a more diagonal structure that may be inverted with improved numerical stability.
Another possible refinement is to employ alternative spectral bases for the radial coordinate.
For example, generalized Laguerre polynomials, whose weight function includes an exponential factor, could be well suited to capture the stiffness induced by the asymptotic decay $e^{-\mu r}$.
Such a basis may provide a more efficient representation of massive scalar fields, especially at larger values of $\mu$.
We leave these explorations to future work to future work.

\appendix

\begin{acknowledgments}
\vspace{0.2cm}
\noindent 
The author acknowledges the Herchel Smith Fellowship at the University of Cambridge for full support, and the College Research Associateship at St John's College, Cambridge for partial support of this work. 
A.K.W.C would like acknowledge Kelvin K.H. Lam for insightful discussion about the work, and sharing the scripts used to produce some of the plots in the paper.  
A.K.W.C would like to thank Pablo Cano, Vitor Cardoso, Richard Dyer, Kelvin K.H. Lam and Nicolas Yunes for their feedback on the initial version of the manuscript. 
The calculations and results reported in this paper were produced using the computational resources of Scientific and High Performance Computing at DAMTP, Cambridge.

\end{acknowledgments}

\section*{DATA AVAILABILITY}

The scalar-field and spacetime-modification solutions constructed in this work are publicly available at \cite{chung_2026_20022233} as Mathematica files.

\bibliography{ref}

\end{document}